\UseRawInputEncoding

\documentclass[twocolumn, twocolappendix]{aastex701}
\usepackage{graphicx} 
\usepackage[dvipsnames]{xcolor}  
\usepackage{booktabs}
\usepackage{amsmath}
\usepackage{bm}
\usepackage{hyperref}
\usepackage{subcaption} 

\usepackage{yfonts}
\usepackage{caption}

\usepackage{ulem}
\definecolor{meridithgreen}{RGB}{0, 140, 0}

\definecolor{christianred}{RGB}{213,94,0}

\def\mesa{\texttt{MESA}} 

\def\sun{$_\odot$}
\def\mlt{$\alpha_\text{MLT}$}
\def\gaia{\textit{Gaia}}
\def\bolddydz{\bm{$\frac{\Delta \rm Y}{\Delta \rm Z}$}}
\def\dydz{$\frac{\Delta \rm Y}{\Delta \rm Z}$}

\begin{document}

\title{Sherlock 'Chrones: The Curious Case of NGC 6791 and Asteroseismic Targets Therein\\Reveals Mixing Length Dependence on Evolutionary Stage}

\author[orcid=0009-0009-7053-6445,sname='Frankel']{Eliza Frankel}
\affiliation{Department of Physics and Astronomy, University of Wyoming, 1000 E. University Avenue, Laramie, WY 82071, USA}
\email{efrankel@uwyo.edu}

\author[0000-0002-8717-127X]{Meridith Joyce}
\affiliation{School of Computing, University of Wyoming, 1000 E. University Avenue, Laramie, WY 82071, USA}
\affiliation{Department of Physics and Astronomy, University of Wyoming, 1000 E. University Avenue, Laramie, WY 82071, USA}
\affiliation{School of Physics and Astronomy, Rochester Institute of Technology, 1 Lomb Memorial Drive, Rochester, NY 14623, USA}
\email[hide]{mxjsps@rit.edu}

\author[0000-0002-4818-7885]{Jamie Tayar}
\affiliation{Department of Astronomy, University of Florida, Gainesville, FL 32611, USA}
\email{jtayar@ufl.edu}

\author[orcid=0009-0003-4467-7718]{Ily Rubio}
\affiliation{School of Computing, University of Wyoming, 1000 E. University Avenue, Laramie, WY 82071, USA}
\email[hide]{}

\author[0000-0002-8878-3315]{Christian I. Johnson}
\affiliation{Space Telescope Science Institute
3700 San Martin Drive
Baltimore, MD 21218, USA}
\email{}

\begin{abstract}
As single--aged populations containing stars spanning a range of masses but otherwise with similar properties, star clusters provide valuable testbeds for stellar evolution theory. The old, metal--rich, open cluster NGC 6791 is well studied, but the availability of asteroseismic observations of individual cluster members allows for a more in--depth analysis. Driven by mounting evidence that the solar assumption for convection used in almost all 1D stellar models is not adequate, and in light of the increased precision provided by modern data and asteroseismic constraints, we built and compared custom isochrones with varying values of the convective mixing length, \mlt, to two clusters with well--determined ages: NGC 6791 and the well--studied, metal--poor globular cluster M92. For both clusters, we found that the best-fitting mixing length parameter decreases monotonically as a function of evolutionary phase, finding values between $0.98-1.10\alpha_{\odot}$ for NGC 6791 and $1.16-1.21\alpha_{\odot}$ for M92. 
In NGC 6791, this corresponds to an 18\% decrease in the best fitting mixing length between the main sequence turn-off and upper end of the red giant branch. The degree of variation found within this one object is therefore comparable to variation across the literature in \mlt\ fits for other non-solar targets, calling single-valued \mlt\ fits to clusters into question. It is also consistent with $\log g$-\mlt\ trends predicted by higher-fidelity, 3D convection analyses.
Our results therefore illustrate the need for more flexibility in our treatment of the mixing length in isochrones to properly model clusters.
\end{abstract}

\section{Introduction} 
\label{sec:Intro}

Star clusters are groups of stars with similar chemical compositions that can be categorized into two major groups: open and globular clusters. Open clusters (OCs) are typically younger ($\lesssim 1$ Gyr), less massive ($\lesssim 10^{5} M_\odot$), and located in the disk of the Milky Way (MW). OCs are mostly comprised of single populations where age spreads are minimal and all stars have approximately the same chemical compositions, except for differences induced by stellar evolution \citep[e.g.,][]{krause_physics_2020}. Globular clusters (GCs) are typically older, having formed in the early universe (e.g., $\sim$13 billion years ago), denser, more massive ($\gtrsim 10^{4} M_\odot$), and located predominantly in the MW bulge and halo \citep{gratton_abundance_2004, krause_physics_2020}. Almost all GCs are observed to have multiple populations, meaning light element chemical abundances are non-uniform among the cluster stars \citep[e.g.,][]{bastian_multiple_2018}. 

Clusters of both kinds host stars at different phases of evolution, creating the perfect laboratory for exploring stellar evolution and mass distributions. For example, the main sequence turn-off (MSTO), or the end of the central H-burning phase in low--mass stars \citep{salaris_evolution_2005}, has a well characterized relationship between age and luminosity for low--mass stars. Alongside the distance to the cluster, the absolute magnitude of the MSTO can be used to determine cluster ages \citep{chaboyer_accurate_1996}. The subgiant branch (SGB) can also be used for age determinations. Luminosities on the SGB scale with core mass, which subsequently scales with MS lifetime \citep{salaris_evolution_2005, kippenhahn_stellar_2012, nataf_accurate_2024}. With measurements of stellar luminosities on the subgiant branch, we are able to determine cluster ages \citep{xiang_time-resolved_2022}. 

The red giant branch (RGB) is another evolutionary feature of clusters that encodes important physical information. This includes the RGB bump (see \citealt{cassisi_critical_1997, cassisi_magnitude_2011, salaris_red_2002, riello_red_2003} for discussion on the importance of the RGB bump), the tip of the RGB (TRGB), and the shape of the RGB. The TRGB occurs as low--mass stars reach the helium flash \citep{salaris_tip_1997, salaris_new_1998}, where the luminosity relies on He core mass (and thus stellar mass).
Additionally, the bolometric magnitude of the TRGB (which comes from models) can be used as a standard candle for accurate distance measurements \citep[e.g., ][]{jacobs_extragalactic_2009, anand_tip_2024, newman_empirical_2024}, making it an important diagnostic of evolutionary behavior \citep{salaris_red_2002, durbin_mcr-trgb_2020}. We can also use the shape of the RGB in broad-band to derive cluster metallicity \citep{sarajedini_technique_1994}.

To explore stellar evolution in clusters, we use isochrones: single--aged and chemically homogeneous models connecting stars of different masses with the same age \citep[e.g., ][]{demarque_y2_2004, dotter_dartmouth_2008, dotter_mesa_2016, pietrinferni_updated_2024, nguyen_parsec_2025}. Isochrones show a snapshot in time of stellar populations, illustrating how stars of different masses reach different evolutionary phases at the same age. We can compare isochrones to clusters with well--determined ages to learn more about the underlying physical processes of stars. However, as with all stellar models, there are many uncertainties that can impact this including: metallicity, $\alpha$--abundance, Helium diffusion, the treatment of convection and choice of convective mixing length, opacity and more \citep{chaboyer_absolute_1995, vandenberg_age_1996, ying_absolute_2025, dufresne_absolute_2026}. To mitigate these, we must be intentional about the input parameters and physical assumptions governing our models. Initial mass and chemical composition are necessary specifications for a stellar track, but they are not the only critical input parameters; the treatment of convection is a necessary consideration.

Modeling convection is notoriously difficult and can significantly impact stellar age determinations. Convection is a 3D process; however, it is necessary to model in 1D with numerous simplifications in order to model stars on evolutionary timescales. 1D convection is parameterized using a diffusive approximation, is static, and does not by itself account for boundary mixing, including convective overshooting into stable, radiative zones \citep{renzini_embarrassments_1987, grossman_theory_1993, nordlund_3-d_2000, anders_convective_2023}.

\begin{deluxetable*}{cc|c|c|c|c}[!]
    \tablecaption{Cluster properties for NGC 6791 and M92.\label{tab:cluster_properties}}
    
        \tablehead{
        & &
        \multicolumn{2}{c|}{\textbf{NGC 6791}} &
        \multicolumn{2}{c}{\textbf{M92}}
        }
        \startdata
        Age & (Gyr) & $8.3 \pm 0.3$ & (1) & $13.80 \pm 0.75$ & (8)  \\
        $\rm{[Fe/H]}$ & & $+0.31 \pm 0.04$ & (2) & $-2.30 \pm 0.10$ & (9, 10, 11)  \\
        $\rm{[\alpha/Fe]}$ & & $+0.06 \pm 0.05$ & (3) & $+0.3$ & (12)  \\
        Z\tablenotemark{a} & & 0.0345 & & 0.000145 & \\
        \bolddydz & & $1.4 \pm 0.1$ & (1) & 1.6 & (13) \\
        RA & (deg)  & 290.220 & (4) & 259.281 & (14)  \\
        Dec & (deg) & 37.778 & (4) & 43.136 & (14)  \\
        $E(B-V)$  & & $0.14 \pm 0.02$ & (1) & 0.02 & (15) \\
        d  & (kpc) & $4.01 \pm 0.14$ & (5) & $8.501 \pm 0.07$ & (16) \\
        $r_\text{\textbf{hl}}$ & (arcmin) & $4.10 \pm 0.16$ & (6) & 1.02 & (17)  \\
        $r_\text{\textbf{hl}}$ & (deg) & $0.06833 \pm 0.00267$ & & 0.017 & \\
        Parallax  & (mas) & $0.192 \pm 0.002$ & (7) & $0.112 \pm 0.010$ & (18)\\
        \enddata
        \tablenotetext{a}{Converted from [Fe/H] \& [$\alpha$/Fe] to Z with a standard conversion assuming the \citet{grevesse_standard_1998} individual element abundances for the Sun, found at \url{https://github.com/mjoyceGR/abundance_converter}}
        \tablecomments{The sources for each table component are as follows: (1) \citet{brogaard_age_2012}, (2) \citet{myers_open_2022}, (3) \citet{villanova_ngc_2018}, (4) \citet{gaia_collaboration_gaia_2023}, (5) \citet{brogaard_age_2011}, (6) \citet{kamann_linking_2019}, (7) \citet{cantat-gaudin_clusters_2020}, (8) \citet{ying_absolute_2023}, (9) \citet{kraft_globular_2003}, (10) \citet{carretta_intrinsic_2009}, (11) \citet{cohen_no_2011}, (12) \citet{vandenberg_models_2000}, (13) \citet{balser_chemical_2006}, (14) \citet{goldsbury_acs_2010}, (15) \citet{pont_hipparcos_1998}, (16) \citet{baumgardt_accurate_2021}, (17) \citet{harris_new_2010}, and (18) \citet{vasiliev_gaia_2021}}

\end{deluxetable*}

\begin{deluxetable*}{l|cc}
    \tablecaption{Input model physics for \mesa\ grid. The same physics is used for both NGC 6791 and M92.\tablenotemark{a} \label{tab:input_params}}

        \tablehead{
        \colhead{\textbf{Parameter}} &
        \colhead{ {\tt MESA} \textbf{Value}} &
        \colhead{\textbf{Model Physics}}
        }
        
        \startdata
        Atmosphere & `{\tt Photosphere}' & PHOENIX, \citealt{hauschildt_nextgen_1999, hauschildt_nextgen_1999-1} \\
        [-0.4ex]
        & & and \citealt{castelli_new_2003} \\
        Convective overshoot & $-$ & No \\
        Diffusion & $-$ & No  \\
        Equation of State & {\tt use\_Skye = .true.} & Skye \citep{jermyn_skye_2021}, \\
        [-0.4ex]
        & {\tt use\_FreeEOS = .true.} & FreeEOS \citep{irwin_freeeos_2004, irwin_freeeos_2004-1,irwin_freeeos_2005, irwin_freeeos_2005-1, irwin_freeeos_2007}, \\
        [-0.4ex]
        & {\tt use\_OPAL\_SCVH = .true.}
        & OPAL \citep{rogers_updated_2002}, \\
        [-0.4ex]
        & & SCVH \citep{saumon_equation_1995}, \\
        [-0.4ex]
        & & \& HELM \citep{timmes_accuracy_2000} \\
        High-temperature opacities\tablenotemark{b}& `{\tt gs98}' & \citet{grevesse_standard_1998} \\
        Low-temperature opacities\tablenotemark{b} & `{\tt lowT\_fa05\_gs98}' & \citet{ferguson_low-temperature_2005} \\
        Mixing Length Parameter & {\tt mixing\_length\_alpha} & 1.0 - 2.5, $\delta_{\alpha_{\rm MLT}} = 0.1$ \\
        Mixing Length Scheme & `{\tt Henyey}' & \citet{henyey_studies_1965}  \\
        Nuclear Reaction Rates & `{\tt basic.net}' & Includes PP chain, CNO cycle, \& helium burning \\
        RGB cool wind scheme & `{\tt Reimers}' & \citet{reimers_circumstellar_1975} \\
        Reimers scaling factor & `{\tt 0.1d0}' & \citet{howell_tess_2026} \\ 
        Solar chemical composition & `{\tt gs98}' & \citet{grevesse_standard_1998}\\
        \hline
        & \textbf{NGC 6791} & \textbf{M92} \\
        [+0.4ex]
        \hline
        [Fe/H] & $+0.31 \pm 0.04$ & $-2.30 \pm 0.10$ \\
        $[\alpha/\text{Fe}]$ & $0.06 \pm 0.05$ & 0.3 \\
        Initial Z & 0.0345 & 0.000145 \\
        Initial Y & 0.2963 & 0.2482 \\
        Masses & 0.80 - 1.17 M\sun & 0.70 - 0.80 M\sun \\
        \hline
        Solar \mlt & \multicolumn{2}{c}{1.72928}\\
        \enddata

\tablenotetext{a}{The \mesa\ inlists are available at both \dataset[doi:10.5281/zenodo.24092924]{https://doi.org/10.5281/zenodo.24092924} and \url{https://github.com/eliza-frankel/sherlock_chrones}}
\tablenotetext{b}{We note that standard opacities without $\alpha$-enhancement were used for both clusters. See Section \ref{subsec:modeling_assumptions} for further discussion.}
\end{deluxetable*}

In our models, we use mixing length theory (MLT), the standard 1D parameterization of convection \citep{bohm-vitense_uber_1958}. Mixing length theory is used to characterize energy transport in convectively inefficient regions of stellar models, or regions where energy flux is partially carried by both convection and radiation. This is the case in superadiabatic regions\footnote{Superadiabatic regions occur when the temperature gradients allow high temperatures to correspond with lower densities, causing the hot gas to rise \citep{nomoto_mixing_1972, joyce_review_2023}} of low--mass stars hosting convective envelopes. MLT posits that in such convection zones, there are cells (or ``parcels") of fluid in pressure but not thermal equilibrium with the surrounding gas. Due to buoyancy forces, hot parcels rise and expand towards the cooler convective regions, while cooler parcels shrink and compress towards the hotter gas. The convective flux transported by these cells, $F_{conv}$, is proportional to the mixing length parameter, \mlt, a measure of convective efficiency \citep{joyce_review_2023}. 
It is a nonphysical, free parameter that changes depending on which stellar evolution code is used and which physical assumptions (e.g., for atmospheric boundary conditions) are adopted, and thus it must be calibrated on a case-by-case basis \citep{nordlund_3-d_2000, joyce_review_2023}.

Altering the value of the mixing length parameter can drastically change a stellar model's main sequence lifetime, model--based age inferences, and more \citep{joyce_not_2018}. Figure 12 of \citet{ying_absolute_2025} shows both observational and theoretical (modeling) considerations that increase error in age determinations, with \mlt\ being one of the key sources of intrinsic model uncertainty that is not accounted for in any publicly available isochrone database.\footnote{While many isochrone and stellar modeling databases allow for variations in, e.g., helium abundance (see BaSTI \citep{hidalgo_updated_2018, pietrinferni_updated_2021, salaris_updated_2021, pietrinferni_updated_2024} \& PARSEC \citep{bressan_parsec_2012, nguyen_parsec_2022, nguyen_parsec_2025}), none allow for variation in $\alpha_\text{MLT}$} For example, when \mlt\ is decreased, it increases the age at which a model will reach a certain temperature and luminosity, prolonging the lifetime of the main sequence \citep{joyce_not_2018, joyce_review_2023}. As a result, varying \mlt\ in the underlying stellar tracks changes the morphology of the SGB and RGB, which impacts age determinations of observed clusters.

Historically, a solar calibrated mixing length has been used when modeling non--solar stars without much physical reason beyond observational limitation. The mixing length parameter likely scales with a number of physical parameters, including metallicity \citep{salaris_new_1996, tayar_correlation_2017}, surface gravity ($\log g$), and effective temperature ($T_{\rm eff}$) \citep{bonaca_calibrating_2012, magic_stagger-grid_2015, viani_investigating_2018, song_impact_2020}. Additionally, there is evidence that \mlt\ should vary as a function of stellar age, with older, cooler stars impacted by convection much more than younger stars \citep{cassisi_galactic_1999}. This means that stars with different $\log g$, metallicites, $T_{\rm eff}$ and ages as the Sun should logically be better modeled by a different \mlt. 

With this in mind, there have been numerous studies (e.g. \citealt{bonaca_calibrating_2012, joyce_not_2018, woody_classical_2026}) which have found that solar calibrated values of \mlt\ are unable to reproduce observed stellar characteristics. Most notably is the well observed and nearby system $\alpha$ Centauri A \& B \citep{guenther__2000, joyce_classically_2018, spada_testing_2019}. Both $\alpha$ Cen A \& B are metal--rich \citep{porto_de_mello_alpha_2008, morel_chemical_2018}; however to replicate observations, $\alpha$ Cen A always requires a sub--solar mixing length parameter while $\alpha$ Cen B does not. Another famous case is HD 140283, sometimes known as the Methuselah star, a metal--poor subgiant star with virtually no reddening and a well known, highly metal-depleted chemical composition. This star has had its age determined to be as old as 15 Gyr \citep{guillaume_age_2024}. However, using sub--solar values for \mlt,  \citet{creevey_benchmark_2015} and \citet{tang_revised_2021} were able to find ages younger than the age of the universe. Additionally, in a study by \citet{bonaca_calibrating_2012}, it was found that when using a solar calibrated mixing length parameter, over 50\% of the best fitting models have an initial helium abundance lower than the primordial value. The most likely cause for these non-physical helium results is using an incorrect choice of $\alpha_\text{MLT}$, as the mixing length and helium abundance are highly degenerate and covariant. It is thus reasonable to draw the conclusion that the solar prescription is not valid in these stars, either.

Concurrently, one of the best methods for learning about individual stars is asteroseismology, or the study of stellar pulsations \citep{aerts_probing_2021, pinsonneault_apokasc-3_2025}. It can be used to determine precise measurements of stellar age, mass, and radius \citep{bellinger_stellar_2019, aerts_probing_2021} while testing stellar interiors and constraining the efficiency of convective mixing both inside and outside of convective regions (e.g., \citealt{chaplin_asteroseismology_2013}). Though there are discrepancies among asteroseismic age determinations (\citealt{tayar_star-crossed_2025}; for NGC 6791, see also \citealt{dufresne_absolute_2026}), asteroseismology is able to estimate cluster properties very precisely \citep{chaplin_asteroseismology_2013}. Seismically determined values like mass, age, and radius can be compared to independent measurements, for example masses found from isochrones, to better age-date coeval stellar populations and constrain stellar parameters \citep{howell_tess_2026}. Additionally, using these seismic determinations of mass and radius and known values of effective temperature and metallicity, asteroseismic targets can be used to constrain the value of \mlt\ \citep{bonaca_calibrating_2012}.

Clusters with reliable and precise age determinations are a great way to calibrate stellar models. With the goal of calibrating \mlt, we chose two well--studied clusters: NGC 6791 and M92. NGC 6791 is an old ($\sim 8.3 \pm 0.3$ Gyr; \citealt{brogaard_age_2012}), metal--rich ([Fe/H] = $+0.31 \pm 0.04$; \citealt{myers_open_2022}) open cluster located $\sim$8 kpc from the Galactic Center and $\sim$1 kpc above the Galactic plane \citep{king_color-magnitude_2005, jilkova_origin_2012}. In addition to having a well-determined age and a non-solar metallicity, asteroseismic data are also available. As a result, it is an ideal candidate for \mlt\ calibration in a non-solar environment.

Meanwhile, M92 is an extremely old ($\sim 13.8 \pm 0.75$ Gyr) and metal--poor ([Fe/H] = $-2.30 \pm 0.1$) globular cluster in the Milky Way's halo \citep[e.g.,][]{cohen_abundances_1979, sneden_oxygen_1991, kraft_globular_2003, cohen_no_2011, ying_absolute_2023}. Numerous previous studies have found a sub--solar mixing length parameter to provide the best fit for M92. For example, using the best fitting \mlt--[Fe/H] relation from \citet{tayar_correlation_2017}, we found the mixing length parameter of M92 to be $\sim$10$\%$ lower than $\alpha_{\odot, \rm{YREC}}$\footnote{i.e., the solar calibrated value of $\alpha_\text{MLT}$ found using the Yale Rotating stellar Evolution Code (YREC; \citealt{pinsonneault_evolutionary_1989, pinsonneault_yrec_2026})}, and \citet{joyce_not_2018} also found a value $\sim$10$\%$ lower than $\alpha_{\odot, \rm{DSEP}}$ using the Dartmouth Stellar Evolution Program (DSEP; \citealt{dotter_dartmouth_2008}). While M92 does not have asteroseismic data available, it is a well--studied globular cluster acting as a validating test case for our isochrone fitting methods.

With accumulating evidence for the necessity of non--solar mixing lengths in models of stars and stellar populations, we generate a bespoke suite of variable--\mlt\ isochrones using Modules for Experiments in Stellar Astrophysics (\mesa ;
\citealt{paxton_modules_2011, paxton_modules_2013, paxton_modules_2015, paxton_modules_2018, paxton_modules_2019, jermyn_modules_2023}) stellar tracks. We use the physical parameters of the metal--rich open cluster NGC 6791 and the metal--poor globular cluster M92 to customize these models. By comparing these isochrones to cluster data and new asteroseismic measurements, we are able to determine the best mixing length parameter. 

In this paper, we show that using a non--solar mixing length parameter for non--solar metallicity populations is the best practice in stellar modeling. Section \ref{sec:observational data} explores our methods of data selection and cleaning. Section \ref{sec:Methods} describes the process of building our isochrones and converting from theoretical to magnitude coordinates. Section \ref{sec:determinations_NGC6791} shows the best mixing length parameters for NGC 6791 using four different methods and Section \ref{sec:M92_compare_to_obs} validates this process with M92. Section \ref{sec:comp_to_3D} compares our results to previous 3D studies while Section \ref{sec:atm_BCs} explores the impact of changing atmospheric boundary conditions. Section \ref{sec:conclusion} outlines our conclusions. Lastly, Section \ref{sec:future_work} discusses the implications for future work.

\begin{figure}
    \centering
    \includegraphics[width=0.93\linewidth]{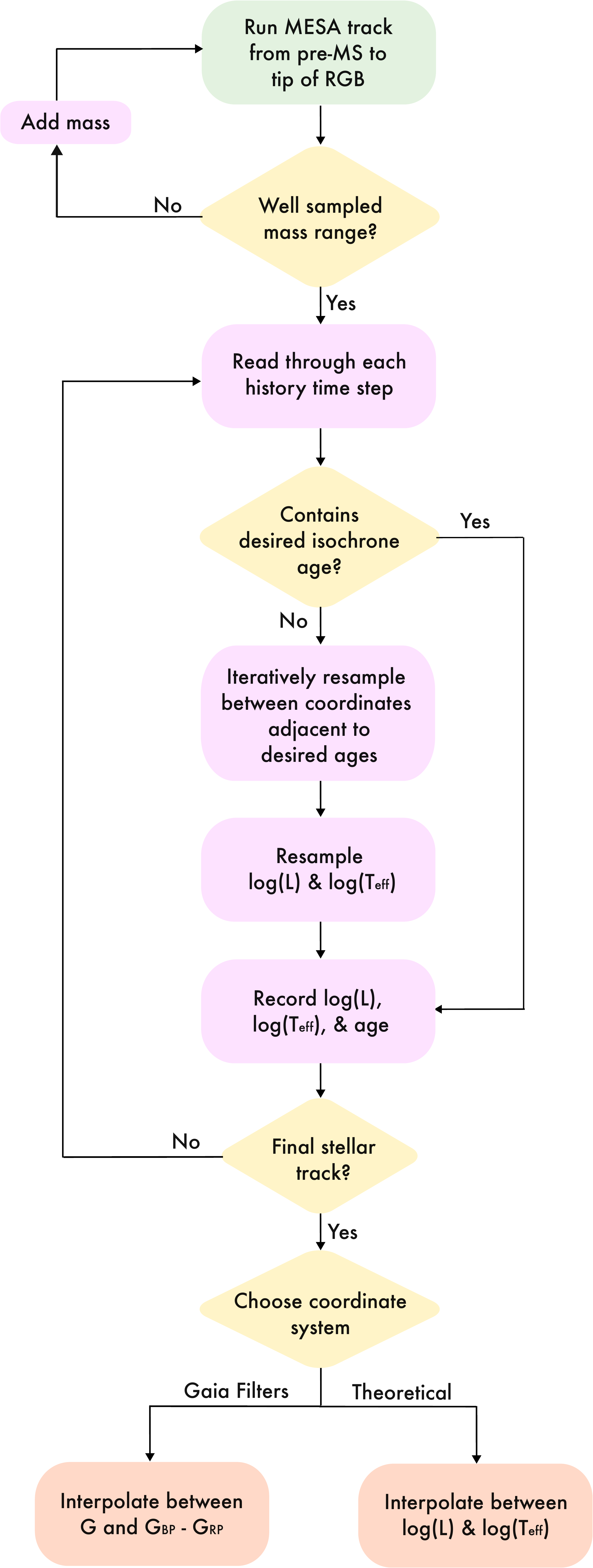}
    \caption{Flow of control for building well sampled, high resolution isochrones. For this study, we used the \textit{Gaia} filter system which is highlighted on the bottom row, however the filters can be replaced with any desired system.}
    \label{fig:flowchart_isochrone}
\end{figure}

\section{Observational Data} 
\label{sec:observational data}
NGC 6791 and M92 are both well-observed clusters with well--known ages and metallicites. Table \ref{tab:cluster_properties} summarizes their physical properties. Because neither cluster has solar metallicity, they are the perfect candidates for exploring the behavior of \mlt\ in non-solar conditions. In our analysis, we used photometric data from \gaia\ DR3 \citep{gaia_collaboration_gaia_2016, gaia_collaboration_gaia_2023-1}. In the case of NGC 6791, these are supplemented by asteroseismic observations from APOKASC-3 \citep{pinsonneault_apokasc-3_2025}.

\subsection{NGC 6791} 
\label{subsec:NGC6791_obs_data}
NGC 6791 is old for an open cluster ($8.3 \pm 0.3$ Gyr; \citealt{brogaard_age_2012}) while also being metal--rich ([Fe/H] = $+0.31 \pm 0.04$; \citealt{myers_open_2022}). To isolate the entire cluster population within Gaia DR3, we extracted stars from the Gaia database that were within 3 half-light radii, $r_\text{hl}$, of the cluster center as described by \citet{vasiliev_gaia_2021}. To clean our data, we followed \citet{vasiliev_gaia_2021}, \citet{riello_gaia_2021}, and \citet{fabricius_gaia_2021}. 
We limited our search to include bright sources ({\tt phot\_bp\_mean\_mag $< 20.3$}), stars at the same distances ({\tt parallax $\pm\ 0.1$}), sources with many visibility periods ({\tt visibility\_periods\_used $> 10$}), sources with consistent $G$-band, BP, and RP photometry ({\tt phot\_bp\_rp\_excess\_factor}), and to exclude probable binary sources ({\tt ruwe $< 1.15$}, {\tt astrometric\_excess\_noise\_sig $< 2$}, and {\tt ipd\_frac\_multi\_peak $< 2$}). The full query with more explanation for each parameter can be found in Appendix \ref{app:query_params_NGC6791}. We then crossmatched the 20 APOKASC-3 asteroseismic targets with the \gaia\ data and have highlighted them in the cluster data (for more information, see Section \ref{subsec:ngc6791_asteroseismic_targets}).

We take an age of $8.3 \pm 0.3$ Gyr from \citet{brogaard_age_2012}, found by comparing eclipsing binaries to cluster CMDs and isochrones. The value of [Fe/H] = $+0.31 \pm 0.04$ \citep{myers_open_2022} came from the OCCAM DR17 \citep{majewski_apache_2017} pipelines, and [$\alpha$/Fe] = $0.06 \pm 0.05$ was found through both the equivalent width and spectrum synthesis methods \citep{villanova_ngc_2018}. \citet{brogaard_age_2012} found $E(B-V)$ by fitting isochrones to a CMD in the $I - V$ plane. To calculate distance, \citet{brogaard_age_2011} used radii, $T_\text{eff}$, and $E(B-V)$ of eclipsing binaries to find the true distance modulus and distance. To determine the half light radius, $r_\text{hl} = 4.10 \pm 0.16$ arcmin, \citet{kamann_linking_2019} used radial star count profiles found by fitting multiple Gaussians using \citet{cappellari_efficient_2002}.

To convert from [Fe/H] and [$\alpha$/Fe] to Z, we used the \citet{grevesse_standard_1998} abundances using a standard conversion.\footnote{The conversion script can be found at \url{https://github.com/mjoyceGR/abundance_converter}.} We adopted the helium enrichment law, 
\begin{equation} \label{eqn:helium_enrich_law}
    \rm Y = \rm Y_{\rm P} + \frac{\Delta \rm Y}{\Delta \rm Z} \times \rm Z,
\end{equation}
where $\text{Y}_\text{P}$ is the primordial helium and \dydz\ is the helium-to-metal enrichment ratio \citep{joyce_stellar_2024}. We have adopted $\text{Y}_\text{P} = 0.248$ \citep{cassisi_initial_2003, peimbert_revised_2007, cyburt_update_2008} and $\frac{\Delta \rm Y}{\Delta \rm Z} = 1.4$ \citep{brogaard_age_2012}.

\subsection{Asteroseismology} \label{subsec:asteroseismology}
Asteroseismic information is available for a selection of NGC 6791 members thanks to the APOKASC-3 catalog \citep{pinsonneault_apokasc-3_2025}, which provides robust physical parameters including spectroscopic metallicities from SDSS-III and SDSS-IV and global asteroseismic quantities for catalog members from Kepler. With information on effective temperature determined by spectroscopy and the stellar pulsation frequency found using asteroseismology, key stellar characteristics like age, mass, and radius can be determined using well-known asteroseismic scaling relationships \citep{kjeldsen_amplitudes_1995, aerts_asteroseismology_2010, bellinger_stellar_2019, aerts_probing_2021}. Stellar radii and masses can be determined very precisely, reaching near 4\% ($1.8\%$ and $3.8\%$, respectively) \citep{pinsonneault_apokasc-3_2025}.

In this analysis, we used the asteroseismically determined masses and surface gravities from APOKASC-3. The masses were determined with star-by-star correction factors: $f_{\Delta \nu}$, which relates the mean density and the observed large frequency $p$-mode spacing, and $f_{\nu \rm max}$ which calibrates $\nu_{\rm max}$ to fundamental Gaia data as a function of position on the giant branch while remaining independent of cluster stars \citep{pinsonneault_apokasc-3_2025}. These correction factors are functions of the large frequency separation ($\Delta \nu$), the frequency of maximum power ($\nu_{\rm max}$), and T$_{\rm eff}$, and are used to find the stellar mass via the scaling relation described in Eqn.\ 1 of \citet{pinsonneault_apokasc-3_2025} (for more information on the scaling relations, see \citealt{chaplin_asteroseismology_2013}, \citealt{ bellinger_stellar_2019}, and references therein).

\subsection{M92} 
\label{subsec:M92_obs_data}
To isolate M92's cluster population, we begin our query by taking a circle with a radius of approximately 3 times the half-light radius. To clean our data, we looked for sources with {\tt phot\_bp\_mean\_mag $< 20.3$}, within $0.1$ of M92's parallax, and that were identified as cluster members using {\tt pmra} and {\tt pmdec}. The exact query parameters  can be found in Appendix \ref{app:query_params_M92}.

For M92, we used the age $13.80 \pm 0.75$, determined through isochrone fitting done by \citet{ying_absolute_2023}. [Fe/H] = $-2.30 \pm 0.10$ came from high-resolution spectra of giants \citep{kraft_globular_2003, carretta_intrinsic_2009, cohen_no_2011}, while the alpha--abundance, [$\alpha$/Fe], was found to be between $+0.3$ to $+0.4$ based on both spectroscopic determinations of individual alpha elements \citep{roederer_heavy-element_2011} and by comparing upper RGB stars to stellar tracks \citep{vandenberg_models_2000}. Using a variety of distance metrics including surface density profiles from either \textit{Hubble Space Telescope} (HST) or \gaia\ (E)DR3, parallaxes of nearby subdwarfs from \gaia, and more, \citet{baumgardt_accurate_2021} found the distance to M92 to be $d = 8.501 \pm 0.07$ kpc. The half-light radius, $r_\text{hl} = 1.02$ arcmin, was found from \citet{harris_catalog_1996} (2010 edition; compiled as an average of values from \citealt{trager_catalogue_1995} and \citealt{van_den_bergh_diameters_1991}) as the radius enclosing half of the total cluster luminosity.

To convert [Fe/H] and [$\alpha$/Fe] to Z, we used the same standard conversion as in Section \ref{subsec:NGC6791_obs_data}. To find Y, we used Equation \ref{eqn:helium_enrich_law}, but adopted a regionally appropriate value of \dydz $= 1.6$ found by looking at metal-poor \ion{H}{2} regions \citep{balser_chemical_2006}.

\begin{deluxetable*}{cc|cc}
    \tablecaption{Mass grid for single isochrones designed to reproduce specific cluster ages. `Increments per Mass Range' provides the mass ranges for each isochrone, as well as the step size within the specified sub-range. The total number of tracks per grid composing the desired isochrone is listed in the bottom row. For $\alpha_\text{MLT} = 1.80$ and $1.90$, we manually added tracks to the upper RGB, resulting in grids of 482 and 477 tracks, respectively. These coordinates for NGC 6791 are visualized in Figure \ref{fig:dense_grid} with a mixing length of $\alpha_\text{MLT} = 1.60$.\label{tab:mass_sampling}}

        \tablehead{
        \multicolumn{2}{c|}{\textbf{NGC 6791}} & \multicolumn{2}{c}{\textbf{M92}}
         \\
         \hline
         Mass Range (M\sun) & Increments ($\delta$M\sun) & Mass Range (M\sun) & Increments ($\delta$M\sun)
        }
        
        \startdata
        $0.80-1.11$ & 0.01 & $0.60-0.74$ & 0.01 \\
        $1.11 - 1.148$ & 0.001 & $0.74-0.757$ & 0.001 \\
        $1.148 - 1.153$ & 0.0001 & $0.757-0.766$ & 0.0005 \\
        $1.153-1.17$ & 0.00005 & $0.766-0.77$ & 0.00003 \\
        \hline
        \multicolumn{2}{c|}{\textit{Total Tracks:} 460} & \multicolumn{2}{c}{\textit{Total Tracks:} 183} \\
        \enddata
\end{deluxetable*}

\section{Methods} 
\label{sec:Methods}
To build isochrones, we first create a grid of evolutionary tracks using the \mesa\ stellar evolution software suite, stable release version r24.08.1. 
Each grid consists of evolutionary tracks beginning with the pre--main sequence and ending when the star reaches the tip of the red giant branch (RGB), as defined by the onset of the helium flash, described in more detail in Section \ref{subsec:modeling_assumptions}.

\subsection{Modeling Assumptions}
\label{subsec:modeling_assumptions}

To model each cluster, a number of physical assumptions were made, as summarized in Table \ref{tab:input_params}. We assumed an atmospheric boundary condition of {\tt photosphere}, which uses photospheric model atmosphere tables,
\footnote{In \mesa, the photospheric tables assume the N. Grevesse \& A. Noels (\citeyear{grevesse_atomic_1993}, GN93) Solar abundance mixture \citep{paxton_modules_2011}, while we use the N. Grevesse \& A. J. Sauval (\citeyear{grevesse_standard_1998}, GS98) mixture. Both GN93 and GS98 use spectroscopic measurements and match meteoric abundances, but GS98 improved their abundances through through updated empirical photometric models by \citet{grevesse_solar_1999} \citep{basu_helioseismology_2008}.}
relying on the PHOENIX \citep{hauschildt_nextgen_1999, hauschildt_nextgen_1999-1} and \citet{castelli_new_2003} model atmospheres \citep{paxton_modules_2011}. A conservative equation for local specific energy was used via {\tt dedt}. To improve energy conservation, {\tt use\_gold2\_tolerance} is turned on, tightening the convergence tolerance used to achieve hydrostatic equilibrium at each time step \citep{paxton_modules_2019}.
We lowered the limit for change in the integrated power at the stellar surface due to triple-$\alpha$ to {\tt delta\_lgL\_He\_limit $= 0.01$} which ensured that \mesa\ took more steps as the helium-burning luminosity changes. Standard opacities without $\alpha$-enhancement were used for both clusters.\footnote{In Section \ref{sec:atm_BCs}, we find that changing input physics that would likely impact the models more substantially (i.e. atmospheric boundary conditions and the depth they are applied) does not change the conclusions of this study. As a result, we remained consistent with choice of no $\alpha$-enhancement for both clusters.} For the RGB cool wind scheme, we used {\tt Reimers}. The equation for mass loss as given by Reimers \citep{reimers_circumstellar_1975} is proportional to $\eta_R LRM^{-1}$, where $\eta_R$ is the scaling factor. For $\eta_R$, we set {\tt Reimers\_scaling\_factor} = 0.1d0 \citep{howell_tess_2026}.

We evolved stars to the tip of the RGB, just before the onset of the helium flash, as defined by a number of stopping conditions. The stellar track terminated when any of the following conditions were met:
\begin{enumerate}
    \item {\tt he\_core\_mass\_limit = 0.47 M\sun} to account for He core mass at the beginning of the helium flash \citep{iben_single_1984, hansen_stellar_1994} in a typical low--mass range between $0.8-2.3\ \text{M}_{\odot}$ \citep{hansen_stellar_1994, prialnik_introduction_2009};
    
    \item The star reached the helium flash, as defined by the start of He fusion via the triple-$\alpha$ process
    ({\tt stop\_at\_phase\_He\_Burn = .true.});
    
    \item {\tt log\_center\_temp\_upper\_limit = 8} because the triple-$\alpha$ reaction dominates in the core around 10$^{8}$ K \citep{kippenhahn_stellar_2012};
    
    \item {\tt power\_he\_burn\_upper\_limit = 10 L\sun}, to catch and stop models that were not terminated before the He flash \citep{kippenhahn_stellar_2012};
    
    \item The star reaches {\tt max\_age = 20 Gyr}. Stars below $\sim0.7$ M$_{\odot}$ do not undergo the helium flash \citep{prialnik_introduction_2009, kippenhahn_stellar_2012} so we impose an extreme age limit to catch the cases that are not otherwise stopped by helium flash physics; and/or
    
    \item Failing these other conditions, we imposed, as a last resort, an \textit{ad hoc} stopping condition defined as a maximum surface gravity of \\ {\tt log\_g\_lower\_limit = 0.5}.
\end{enumerate}
The inlists used to create the model grids are available at \dataset[doi:10.5281/zenodo.24092924]{https://doi.org/10.5281/zenodo.24092924}.\footnote{The inlists and all other relevant code are also available at \url{https://github.com/eliza-frankel/sherlock_chrones}.}

\subsection{Model Grid} 
\label{subsec:Model Grid}
To build isochrones for both NGC 6791 and M92, we created boutique model grids for each, tailoring the metallicities of the stellar tracks to the literature values for each cluster. We launched a series of stellar tracks between $0.80 - 1.17\ \text{M}_{\odot}$ for NGC 6791 and $0.70 - 0.80\ \text{M}_{\odot}$ for M92. The mass ranges we used for each isochrone varied with the goal of a densely sampled RGB, which resulted in non-uniform spacing in mass coordinates (e.g., greater mass density along the more quickly evolving regions, such as the RGB).

To determine the mass ranges and appropriate mass step sizes, $\delta M$, we ran a small, coarse grid of stellar tracks and decreased $\delta M$ as needed in under-sampled regions until the isochrone features were well resolved and the gaps between adjacent mass coordinates were minimized, as defined in Figure \ref{fig:flowchart_isochrone}. 
We found morphology was best recovered when each isochrone for NGC 6791 and M92 had 460 and 183 tracks, respectively. The values of $\delta M$ and the regimes in which they are used can be found in Table \ref{tab:mass_sampling}. For two values of \mlt\ for NGC 6791 ($\alpha_\text{MLT} = 1.80$ and $1.90$), we manually added masses to remedy gaps on the upper RGB that were not present for other \mlt, resulting in isochrones with 482 and 477 tracks, respectively. 

Critically, we varied the mixing length parameter, \mlt, between $1.0-2.5$ times the pressure scale height, $H_P$, in increments of 0.1 to see how the morphology of the RGB changed in response. The inclusion of varying \mlt\ to our model grid and ultimately our suite of isochrones added another dimension for exploration that is not currently available in any open-source isochrone database.

\subsection{Solar Calibration}
\label{subsec:solar_calibration}
To determine how the best fitting mixing length parameter compares to the solar value, we performed a solar calibration using our input parameters and model physics. We used a genetic algorithm to search this parameter space, following the method in I. Rubio et al. 2026 (\textit{in prep}). Genetic algorithms follow the Darwinism theory of evolution and use mating, mutation, death and preservation to evolve a population of individual models per generation. The free parameters include $\alpha_\text{MLT}$, $Z_\text{in}$,  and $ \frac{\Delta Y}{\Delta Z}$.  We treat $Y_\text{in}$ as a constrained variable using the helium enrichment law (Equation \ref{eqn:helium_enrich_law}).  In contrast, when $Y_\text{in}$ is treated as an independent free parameter, this gives us sub-primordial helium values regardless of the method (an issue others have identified with MESA  \citep{bonaca_calibrating_2012}). Validation for the GA was done using two standard brute force parameter sweeps and costs are include in our computational specifications (Section \ref{subsec:computational specs}).  The brute force parameter sweeps confirmed that a fixed $\frac{\Delta Y}{\Delta Z}=1.4$, borrowing the value used for NGC 6791 isochrone models, consistently found luminosity higher than solar. The second brute force with a varied  $\frac{\Delta Y}{\Delta Z}$ confirmed we were calibrating to non-solar conditions while allowing a preferred value to fall out, typically below 1.5. See forthcoming deepdive in I. Rubio et al. 2026 (\textit{in prep}) for more information.

We found a solar calibrated mixing length parameter of $\alpha_{\odot} =1.72928 $ using the genetic algorithms. Through this calibration, we normalized our results to the solar scale which reduced biases from the input physics and numerical inconsistencies that are propagated into the free parameters.

\subsection{Computational Specifications} \label{subsec:computational specs}

To build these model grids, we utilized the high performance computing resources from the University of Wyoming's Advanced Research Computing Center (ARCC), which hosts 25 CPU nodes with 96 cores and 175 CPU nodes with 32 cores each with an AMD EPYC 9454 processors. Calculations were distributed across 275 nodes using a total of 2200 cores, with 1 thread per core. As a single instance of a \mesa\ evolutionary calculation does not benefit from more than roughly 18 threads, we assigned 14 to a typical calculation. In total, we used approximately 48,500 CPU hours to create a grid of 7399 stellar tracks, spanning 16 distinct alpha MLT values and
the masses described in Table \ref{tab:mass_sampling} for NGC 6791. Roughly 10,000 CPU hours to build a grid of 2928 stellar tracks was required for M92. 

The solar calibrations were assigned 12 cores to a typical \mesa\ evolutionary model utilizing the high performance computing resources from the University of Wyoming's ARCC. Using the genetic algorithm, with a terminal generation of 250 and population of 100 models per generation, resulting in a max job count of 25,000 models, used roughly 155,000 core hours. 
This constitutes a significant CPU cost improvement over brute-force methods, which at final \mlt\ resolution would require running of order 10$^{10}$ models.
In the adapted form of a brute-force grid, where certain regions were sampled at iteratively higher resolution as good solutions were found in those regions, 19,954 core hours were used to generate 52,000 models. For a detailed discussion of the cost-related advantages and disadvantages of this method, see I. Rubio et al. (\textit{in prep)}.

\subsection{Isochrone Building} \label{subsec:isochrone building}

\begin{figure}[t!]
    \centering
    \includegraphics[width=0.9\linewidth]{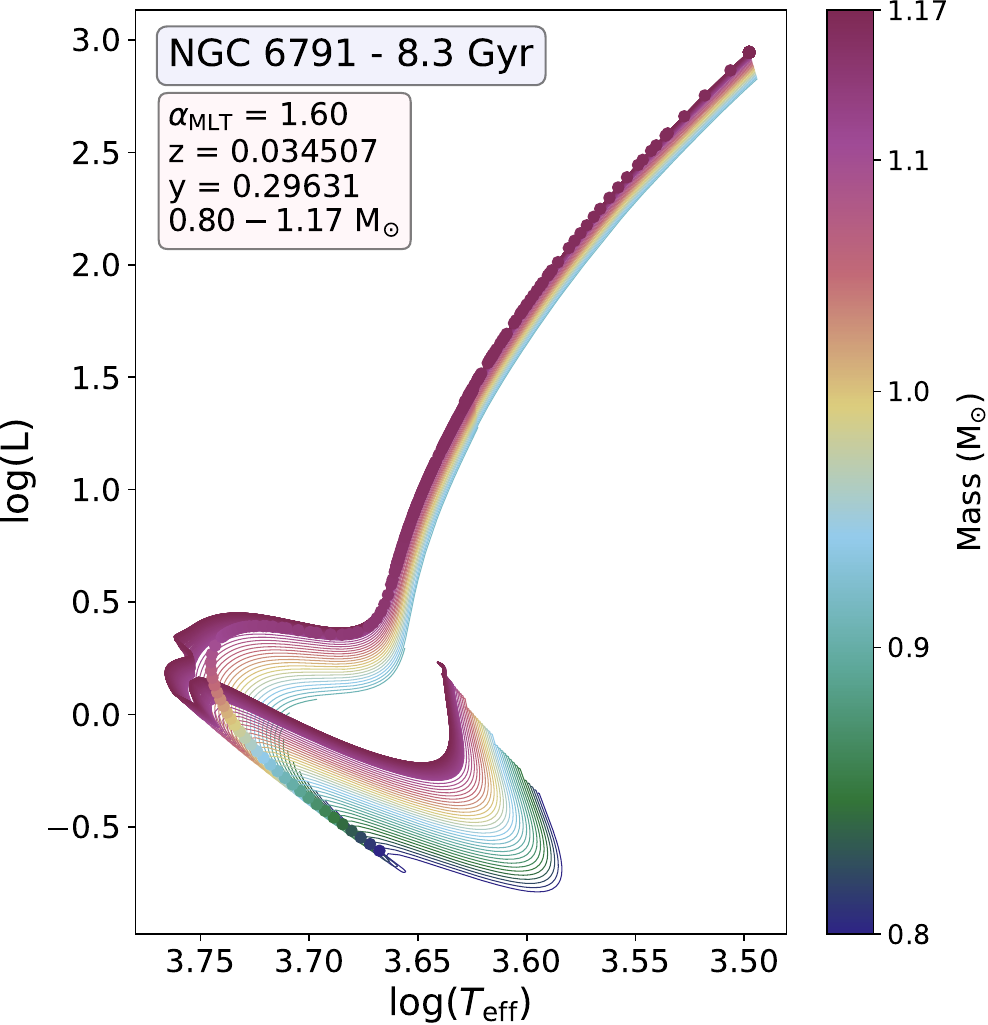}
    \caption{Modeled stellar grid for NGC 6791 with $\alpha_\text{MLT} = 1.60$. The colored lines show the 460 individual tracks between $0.80-1.17$ M\sun, from the pre--MS to the tip of the red giant branch. The dots on each track represent the point closest to 8.3 Gyr, with the color corresponding to the mass of the stellar track, as shown in the colorbar. The density of stellar tracks is important, especially on the RGB as the resulting morphology can significantly change with a less dense sample.}
    \label{fig:dense_grid}
\end{figure}

\begin{figure*}
    \centering
    \includegraphics[width=0.65\linewidth]{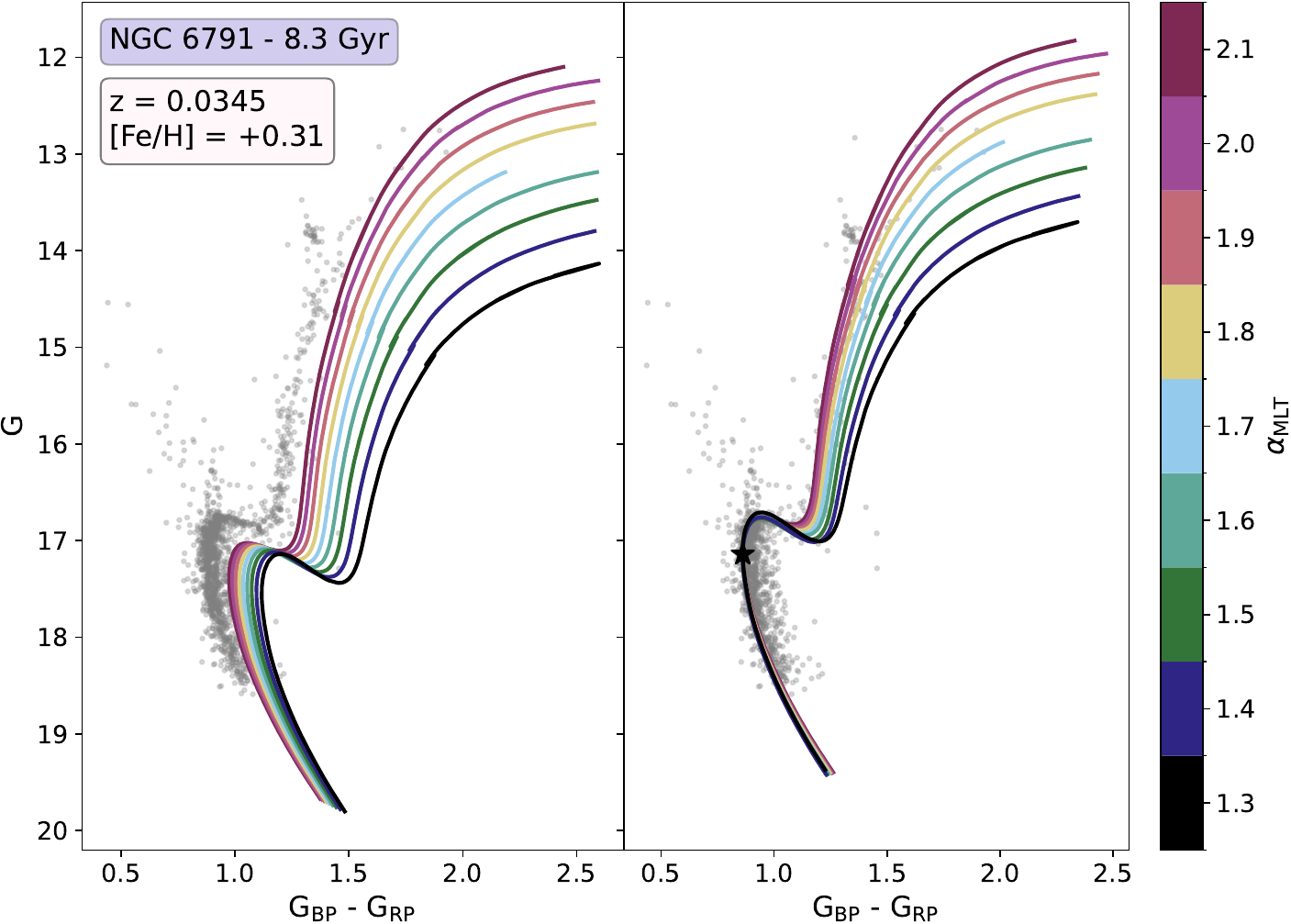}
    \caption{NGC 6791 \gaia\ data with varying \mlt\ isochrones overlaid. Only isochrones with \mlt\ between $1.30 - 2.10$ are included to minimize crowding, but all mixing length parameters between $1.00 - 2.50$ were used in the analysis of this cluster. \textit{Left:} Isochrones on NGC 6791 with no main sequence turn--off pinning. \textit{Right:} Isochrones aligned at the main sequence turn--off, denoted by the black star.}
    \label{fig:NGC6791_vs_all_isochrones}
\end{figure*}

Once a dense grid of stellar tracks had been run to completion, they were used to make isochrones.
The process for building isochrones using \mesa\ stellar tracks is outlined in Figure \ref{fig:flowchart_isochrone}. 
We first found the location in a stellar track's evolution closest to the desired isochrone age to within a tolerance of $10^{-3}$ Gyr. When there are no ages within this tolerance present, we interpolated between the two nearest coordinates for age, mass, effective temperature, luminosity, and surface gravity. This additional interpolation step allowed us to determine the stellar characteristics at the closest possible moment of evolution, without necessarily having a stellar track that has recorded the physical parameters at precisely the desired age coordinate.

Next, we used Piecewise Cubic Hermite Interpolating Polynomial \citep[PCHIP;][]{fritsch_monotone_1980, kahaner_numerical_1989} to interpolate between each stellar track, represented as a mass coordinate on the isochrone. This method allowed us to smoothly connect adjacent mass coordinates without running stellar tracks at prohibitively high densities. To mitigate unintentional biases from this interpolation scheme, we manually created dense grids of \mesa\ stellar tracks as described in Section \ref{subsec:Model Grid}. An example of this can be seen in Figure \ref{fig:dense_grid}, which shows the process of building an 8.3 Gyr isochrone with a mixing length parameter of $\alpha_\text{MLT} = 1.60$. The lines show a total of 460 \mesa\ stellar tracks, colored by mass, while the circular markers represent where the coordinate of the desired isochrone age lands on its associated track. This figure emphasizes the importance of densely sampling throughout all evolutionary stages, especially the RGB: low-mass stars move more quickly through this phase, requiring smaller mass step sizes to properly resolve its morphology. Our maximum resolution of 0.00005 M$_{\odot}$ can be compared to, e.g., the mass resolution of $0.001$ M$_{\odot}$, at which \citet{dotter_mesa_2016} found discontinuities at the transition from radiative to convective cores on the MS.

\subsection{Converting Theoretical Coordinates to Magnitudes} 
\label{subsec:theory_to_mags}

To compare our theoretical isochrones to observational \gaia\ data, we transformed luminosity to \gaia\ $G$-band magnitudes and $T_\text{eff}$ to $G_{\rm BP} - G_{\rm RP}$. The exact steps taken for the conversion between coordinate spaces are described in detail in Appendix \ref{app_subsec:theory_to_mags}.

We use both the modeled luminosity and effective temperature to calculate the absolute magnitudes in both $G$-band and $G_{\rm BP} - G_{\rm RP}$-band (Equations \ref{eqn:L_to_Mbol} - \ref{eqn:Teff_to_Mbol}). This magnitude conversion relies on calculating bolometric corrections (BC) in each filter. To do this, we used the Casagrande and VandenBerg bolometric correction code, which interpolated BCs from stars depending on their effective temperature, surface gravity, metallicity, and reddening \citep{casagrande_synthetic_2014, casagrande_synthetic_2018, casagrande_use_2018}. The BCs were calculated with a solar bolometric magnitude of $M_{bol,\odot}$ = 4.75 \citep{allen_astrophysical_1976, torres_use_2010} and utilizing the MARCS model atmospheres \citep{gustafsson_grid_2008}. Though we used the PHOENIX model atmospheres in our evolutionary calculations, there is a general agreement between the two above 4000 K. Below this, there is some deviation, with the PHOENIX atmospheres appearing slightly warmer than MARCS (see Fig. 19 of \citealt{gustafsson_grid_2008}). Additionally, MARCS utilized the line data from \citet{kurucz_atomic_1995} while PHOENIX utilized the line data from \citet{rothman_hitran_1992} and \citet{kurucz_diatomic_1993}. The majority of the asteroseismic targets for NGC 6791 are at temperatures greater than 4000 K, so we chose to use the Casagrande and VandenBerg code to calculate the BCs despite these differences, and utilized the dust extinction ratio value of $R_{V}$ = 3.1 \citep{cardelli_relationship_1989} in the calculations. 

\section{\mlt\ Determinations for NGC 6791}
\label{sec:determinations_NGC6791}

We shifted our isochrones from theoretical coordinates $T_{\rm eff}$ and $\log L$ to $G$ and $G_{BP} - G_{RP}$, as described in Section \ref{subsec:theory_to_mags}, to directly compare them to the \gaia\ data. To account for E(B-V) and reddening, we applied the \gaia\ (E)DR3 extinction law,\footnote{The (E)DR3 extinction law can be found at \url{https://www.cosmos.esa.int/web/gaia/edr3-extinction-law}} and solved for $A_{m} = k_{m}A_{0}$, where $m$ is each individual \gaia\ filter and $A_{0}$ is the monochromatic extinction at 541.4 nm, for each star in the cluster.

Despite the availability of reddening and extinction information for the majority of targets, there remained an offset between the magnitude-corrected NGC 6791 data and the isochrones in raw theoretical coordinates. To remedy this while avoiding the pitfalls of visual inspection, we devised a method to align the observations and models based on a definition of the main-sequence turn-off.
The MSTO is defined as the bluest point on the MS \citep{chaboyer_accurate_1996}, making it a distinguishable morphological feature.
To determine the MSTO while minimizing any biases that could be imparted by visually choosing a turn--off point, we use the ridgeline method \citep{stetson_homogeneous_2005, zorotovic_globular_2008, alonso-garcia_mapping_2011, simon-zsok_empirical_2025}. A ridgeline is a way of compressing the data to a representative line of mean density following its shape. Once we had a ridgeline, we identified the bluest (hottest) point along it as the MSTO and used this coordinate to represent the cluster's MSTO. As isochrones are already single-valued in $\log g$ or luminosity from the zero age main sequence to the TRGB, identifying an isochrone's MSTO was achieved by selecting the hottest coordinate, which we did self-consistently for each isochrone. The MSTO identified from the cluster ridgeline was then used as the anchor point for the equivalent coordinate on the isochrones, which were shifted into the observational (data) coordinate frame. The right-hand panel of Figure \ref{fig:NGC6791_vs_all_isochrones} demonstrates this alignment. By aligning data with theory using a well-defined reference point, we were able to isolate the impact of the mixing length on the post-MSTO morphology of the isochrones.

\subsection{Asteroseismic Targets} 
\label{subsec:ngc6791_asteroseismic_targets}

\begin{figure}[h!]
    \centering
    \includegraphics[width=0.95\linewidth]{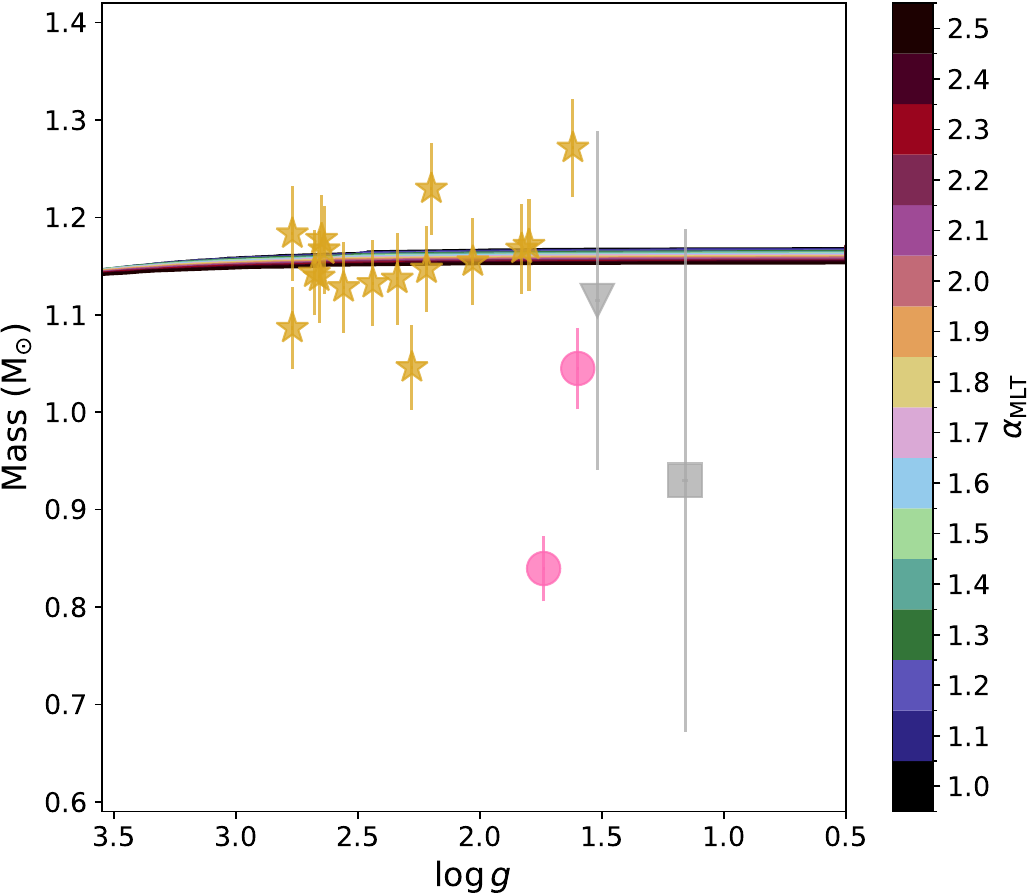}
    \caption{Asteroseismic targets of NGC 6791 on a $\log g$ vs mass plane with uncertainties in both surface gravity and mass, compared to each bespoke isochrone. This plot shows that the choice of \mlt\ does not matter for targets with accurate seismic parameters. The stars are colored by the same classifications as Figure \ref{fig:ngc6791_logg_isochrone}.}
    \label{fig:ngc6791_logg_mass}
\end{figure}

\begin{figure}[!h]
    \centering
    \includegraphics[width=0.9\linewidth]{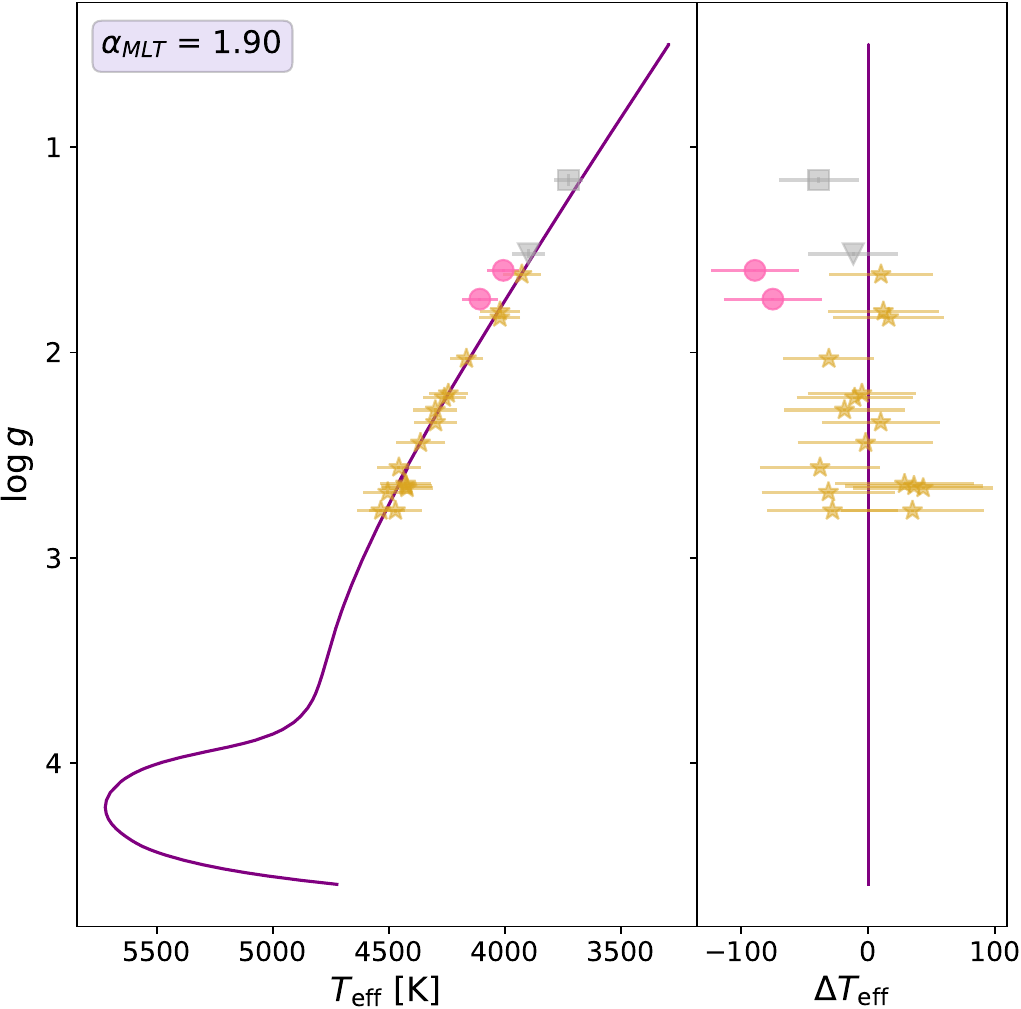}
    \caption{Asteroseismic targets of NGC 6791 on a $\log g$ vs $T_{\rm eff}$ isochrone with $T_{\rm eff}$ residuals and a mixing length parameter of $\alpha_\text{MLT} = 1.90$. The targets are colored by their seismic classifications (Gold and Silver) with two targets suspected to be AGB stars in pink (and thus excluded in isochrone fits). The silver square, KIC 2571093, does not appear in the \gaia\ data, discussed briefly in Section \ref{subsec:ngc6791_asteroseismic_targets}. The uncertainties in the left panel have been multiplied by 2 for visualization purposes; even with the scaling, the uncertainties in $\log g$ remain so small they are not visible. The uncertainties in the right panel have not been scaled.}
    \label{fig:ngc6791_logg_isochrone}
\end{figure}

\begin{figure*}
    \centering
    \includegraphics[width=0.85\textwidth]{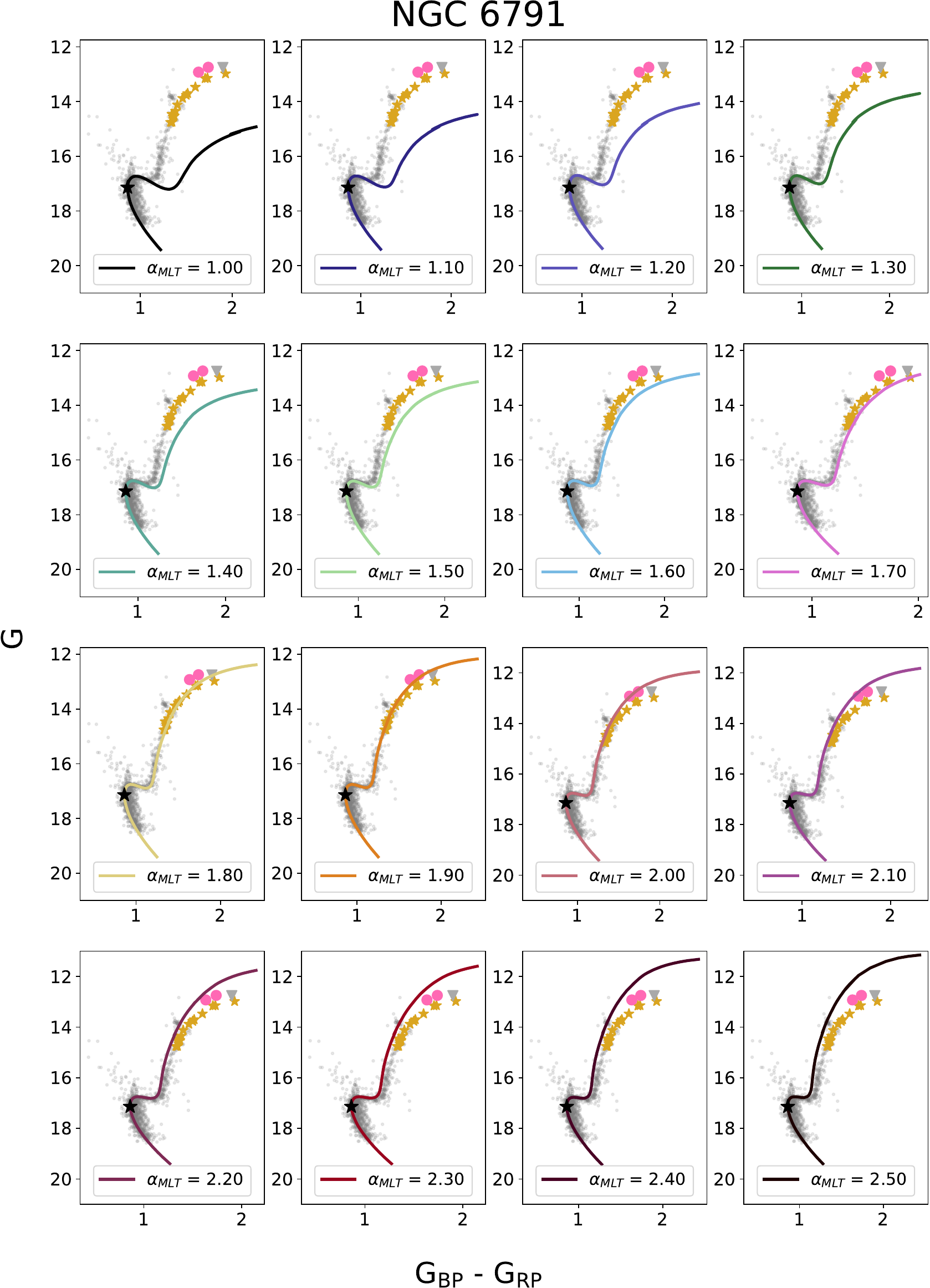}
    \caption{NGC 6791 and the asteroseismic targets compared to each isochrone with varying \mlt\ values. Each isochrone has been shifted slightly to align at the main sequence turn off (denoted by the black star), found using the ridgeline method (Section \ref{sec:determinations_NGC6791}).}
    \label{fig:NGC6791_all_isochrones_grid}
\end{figure*}

For individual stars, asteroseismology provides much more precise measurements of important stellar properties like $\log g$ and mass. In addition to comparing isochrones against the \gaia\ cluster data as a whole, we can compare isochrones to these targets, individual cluster members with asteroseismic measurements, as a way of independently testing the models.

Figures \ref{fig:ngc6791_logg_mass}, \ref{fig:ngc6791_logg_isochrone}, and \ref{fig:NGC6791_all_isochrones_grid} show the asteroseismic targets separated into three categories: yellow stars, grey markers, and pink circles. The yellow stars have the asteroseismic classification of ``Gold'' detections (as defined in \citealt{pinsonneault_apokasc-3_2025}), meaning there are multiple precise measurements of the global asteroseismic characteristics $\nu_{\rm max}$ and $\Delta \nu$ for these targets. The grey markers (KIC 2437496 and KIC 2571093) have the asteroseismic classification of ``Silver'' detections, meaning there are fewer measurements of $\nu_{\rm max}$ and $\Delta \nu$ and there are larger uncertainties. The grey square (KIC 2571093) is an interesting target that does not appear in the \gaia\ query, discussed more in Section \ref{subsec:ngc6791_asteroseismic_targets}.

Lastly, the pink circles (KIC 2437965 and KIC 2569935) are targets we suspect to be on the asymptotic giant branch (AGB). Figure \ref{fig:ngc6791_logg_mass} shows the targets in the mass-$\log g$ plane, where the two pink targets fall significantly below the modeled and observed masses for red giants in this cluster (for more discussion of RGB mass loss in clusters, see e.g., \citealt{miglio_asteroseismology_2012}, \citealt{brogaard_age_2012}, and \citealt{handberg_ngc_2017}). These two stars could be the product of binary interactions. In addition to being less massive than expected if they were on the RGB, these stars are also hotter than other stars at the same metallicity and $\log g$, as shown in Figure \ref{fig:ngc6791_logg_isochrone} \citep{vrard_red_2025} and bluer than the main RGB sequence on the color-magnitude diagram (Figure \ref{fig:NGC6791_all_isochrones_grid}). Because these targets are likely to be AGB stars or otherwise deviate from red giants that evolved as single stars, they are not used when determining the best fitting isochrone.

In Figure \ref{fig:ngc6791_logg_mass}, we show the asteroseismic targets of NGC 6791 compared to all isochrones on the mass-$\log g$ plane, along with uncertainties in both mass and surface gravity from \citet{pinsonneault_apokasc-3_2025}. When comparing to the expected isochrone masses, we see that the majority of the targets are consistent with the predicted masses. Comparing the asteroseismic targets to each isochrone, we see that the spread in mass across isochrones of different mixing lengths is smaller than any given mass uncertainty for the asteroseismic targets. The results from \citet{morales_model_2025} are in agreement; these authors found that with accurate asteroseismic parameters for red giants (in this case, the Gold targets), model choice is less important, with observational uncertainties dominating over theoretical uncertainties.

While many of the targets fall within the expected mass range, there are some outliers. The two possible AGB targets in pink are both below the modeled masses as previously discussed (see \citealt{hekker_asteroseismic_2011, stello_amplitudes_2011, stello_asteroseismic_2011, wu_asteroseismic_2014, tayar_star-crossed_2025, howell_tess_2026}). There are two Gold detections (KIC 2437507 and KIC 2569204) that appear more massive than expected, which could be explained if they are, for example, merger products, or they have otherwise undergone accretion from binary evolution. For more detailed seismic analyses of these targets, see \citet{basu_sounding_2011, hekker_asteroseismic_2011, stello_amplitudes_2011, stello_asteroseismic_2011, wu_asteroseismic_2014, covelo-paz_new_2023, tayar_star-crossed_2025, howell_tess_2026}.

Lastly are the two Silver detections, which both have large mass uncertainties. The most luminous star (KIC 2571093) is much less massive than the models predict. Though a clear conclusion cannot be drawn due to the large uncertainties, one possible explanation is that one or both is an unresolved binary star \citep{evans_evidence_2018, gandhi_astrometric_2022}--as may be suggested by very large {\tt astrometric\_excess\_noise\_sig}---or a possible AGB star that has undergone mass loss \citep{de_marco_dawes_2017} and is at higher-than-expected temperatures \citep{vrard_red_2025}.

\subsection{Methods of Comparison}
\label{subsec:ngc6791_methods}

\begin{figure}[t!]
    \centering
    \includegraphics[width=0.7\linewidth]{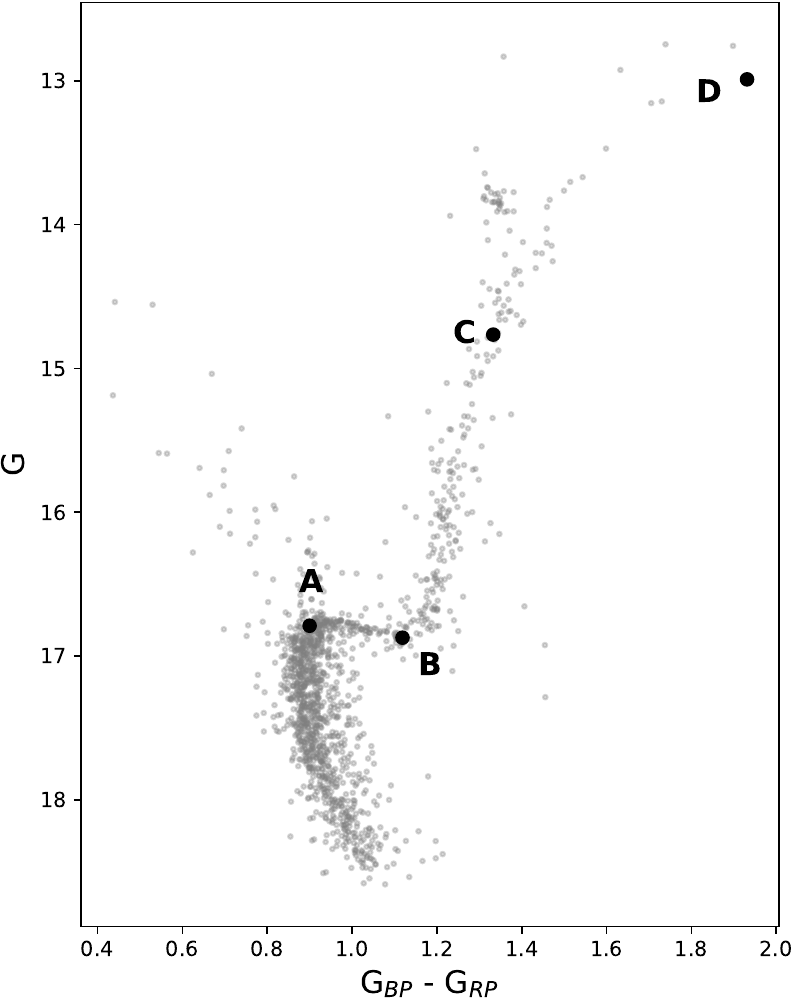}
    \caption{Highlighting four important points in the morphology of NGC 6791. A is the local maxima, or the brightest point before the RGB. B is the local minimum, or the dimmest point after the MSTO. Points C and D are the dimmest and brightest asteroseismic target for NGC 6791.}
    \label{fig:morphology_definitions}
\end{figure}

To find the best convective mixing length parameter, we first compared our customized suite of 8.3 Gyr isochrones with metallicity Z = 0.0345 to the \gaia\ cluster data for NGC 6791. The left panel of Figure \ref{fig:NGC6791_vs_all_isochrones} shows where the isochrones fall in \gaia\ band color-magnitude space, with the best reddening and extinction corrections available for NGC 6791, as compared to the cluster data. For the sake of clarity in Figure \ref{fig:NGC6791_vs_all_isochrones}, we only show isochrones with mixing lengths between $\alpha_\text{MLT} = 1.30 - 2.10$, but all values between $\alpha_\text{MLT} = 1.00 - 2.50$ are used in this analysis. The values of \mlt\ are indicated by color (visualized in the attached colorbar), and are ordered from largest to smallest for the highest to lowest $G$-band magnitude at the TRGB. To find the main sequence turn--off point for alignment, we use the ridgeline method as described in Section \ref{sec:determinations_NGC6791}, aligning the bluest (hottest) point of each isochrone at the same location. 
Results of this method are shown in the right panel of Figure \ref{fig:NGC6791_vs_all_isochrones}, which highlights the changing slope of the RGB as a function of $\alpha_\text{MLT}$ and denotes the alignment point by the black star.

Figure \ref{fig:NGC6791_all_isochrones_grid} shows each individual isochrone of given \mlt\ overlaid on the \gaia\ data for NGC 6791 with the asteroseismic targets highlighted on the RGB. Notably, as we increase the values of \mlt, the slope of the red giant branch becomes steeper, and the subgiant branch becomes flatter and shorter. To find the best-fitting mixing length to NGC 6791 in as quantitative a way as possible, we chose to use four different metrics:
\begin{enumerate}
    \item SGB length, as indicated by the segment between the first local maxima and minima after the MSTO; 

    \item average temperature of the RGB;
    
    \item curvature of the RGB, defined by the slope of a straight line drawn between the least and most luminous asteroseismic targets; and 

    \item offset from brightest asteroseismic target, as a means of exploring consistency around the TRGB specifically. 
\end{enumerate}
To use these metrics, we defined four critical points in the cluster morphology that can be seen in Figure \ref{fig:morphology_definitions}. We define point A as the first local maximum in $G$-band magnitude following the MSTO, point B as the first local minimum in $G$-band following point A, and points C and D as the equivalent G-band coordinates on the cluster data corresponding to the dimmest and brightest asteroseismic targets, respectively. Our goodness-of-fit metrics are subsequently defined using these reference points.

\subsection{Metric 1: Subgiant Branch Length}\label{subsec:ngc6791_sgb}

\begin{figure}
    \centering
    \includegraphics[width=0.75\linewidth]{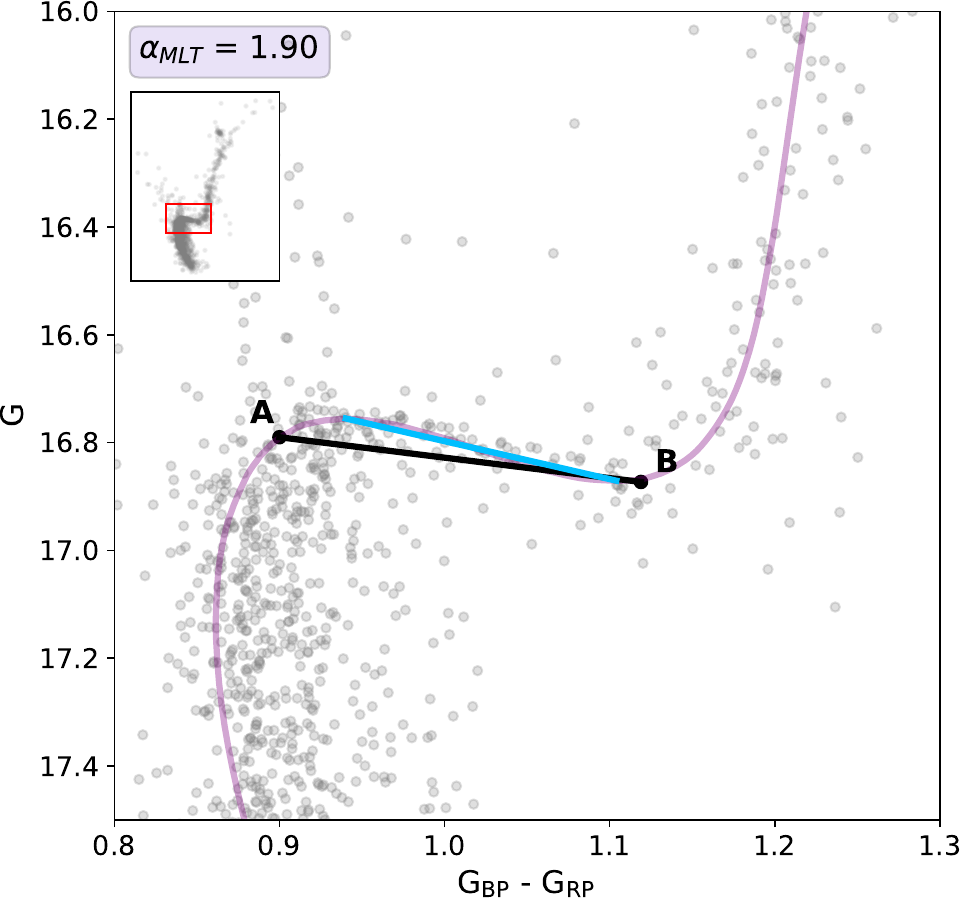}
    \caption{Zoom in on SGB to show comparison of subgiant branch length for the $\alpha_{\rm MLT} = 1.90$ isochrone (blue line) to the subgiant branch of NGC 6791 (black line). The purple line is the entirety of the best fitting isochrone. The inset shows the SGB in relation to the whole cluster.}
    \label{fig:subgiant_length_metric}
\end{figure}

As discussed in Section \ref{sec:Intro}, the lifetime of a stellar model can significantly change depending on the input physics used. For example, when low--mass models have larger centrally mixed regions, the duration of the SGB shortens (see Figure 8 of \citealt{maeder_stellar_1975}) \citep{chaplin_asteroseismology_2013, joyce_not_2018}. On the SGB, stars have a well--defined age-luminosity relationship relating to their core masses, which makes them a sensitive age indicator \citep{salaris_evolution_2005, kippenhahn_stellar_2012, nataf_accurate_2024}. As a result of this relationship, we can compare the length of the SGB to isochrones with varying mixing length parameters to determine which model fits this aspect of the cluster best.

We begin by defining the SGB as the area between the local maxima and minima after the MSTO, or points A and B in Figure \ref{fig:morphology_definitions}. We draw a straight line between these on the cluster data and each isochrone, and calculate the Pythagorean
distance between the points, 
\begin{equation}
\label{eqn:pythag_dist}
    \begin{aligned}
        d & = \sqrt{(x_\text{A,cluster} - x_\text{A,iso})^{2} + (y_\text{A,cluster} - y_\text{A,iso})^{2}} \\
        & + \sqrt{(x_\text{B,cluster} - x_\text{B,iso})^{2} + (y_\text{B,cluster} - y_\text{B,iso})^{2}},
    \end{aligned} 
\end{equation}
where $x$ refers to either $\log T_\text{eff}$ or $G_\text{BP} - G_\text{RP}$ and $y$ refers to either $\log g$ or Gaia $G$-band magnitude, respectively.

The SGB of NGC 6791 is shown as a black line in Figure \ref{fig:subgiant_length_metric} and the SGB of the isochrone is shown as the purple line.
The isochrone with the smallest Pythagorean distance and which thus fits the cluster's subgiant branch best has a mixing length parameter of $\alpha_{\rm MLT} = 1.90$ (or 1.10$\alpha_{\odot}$), which can be seen in Figure \ref{fig:subgiant_length_metric}. While the isochrone's SGB is not a perfect match, when compared to the cluster's SGB, its length is closest.

\subsection{Metric 2: Average Red Giant Branch Temperature} \label{subsec:avg_rgb_temp}

\begin{figure}[t!]
    \centering
    \includegraphics[width=0.85\linewidth]{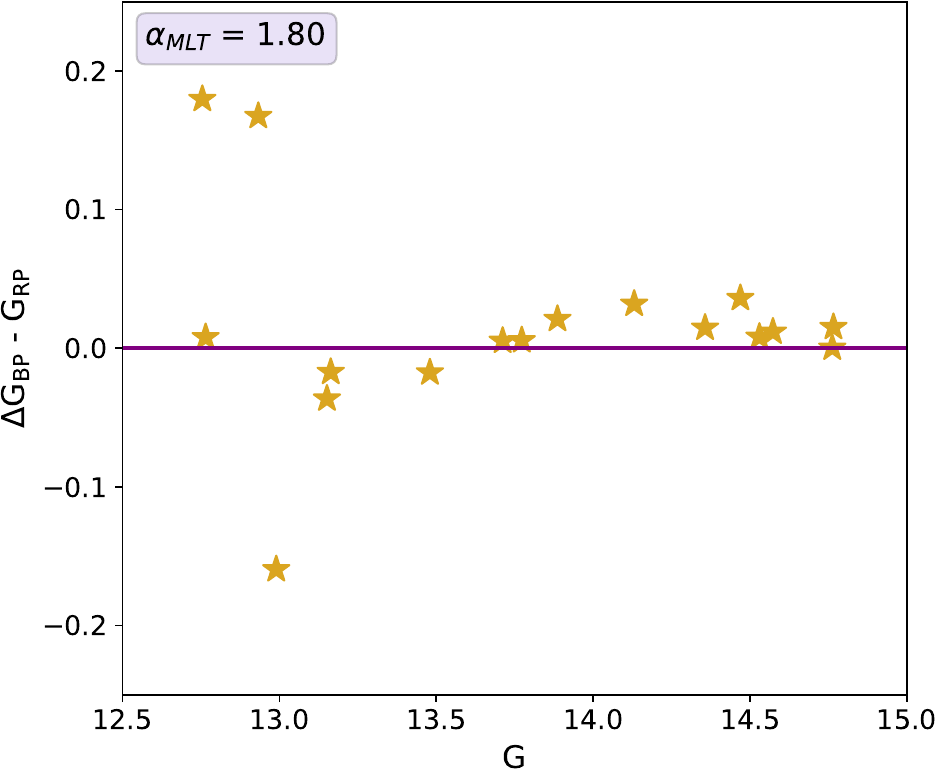}
    \caption{Comparison of the color of the Gold sample asteroseismic targets to the $\alpha_{\rm MLT} = 1.80$ isochrone for NGC 6791.}
    \label{fig:avg_rgb_temp}
\end{figure}

It is known that the slope and temperature of the RGB is sensitive to choice of mixing length parameter \citep[e.g.,][]{ jimenez_stellar_1996, maraston_evolutionary_2005, joyce_review_2023}. With asteroseismic targets providing a means of independent model verification, we can compare temperatures between observations and isochrones, constraining the mixing length. Finding the point on the isochrone with the same $G$ magnitude as the asteroseismic targets, we then determine the difference between their respective $G_{BP} - G_{RP}$ magnitudes. We calculate the 1--dimensional Pythagorean distance between the temperature of the observed and modeled, and find that the best fitting isochrone has a mixing length parameter of \mlt\ = 1.80 ($1.04\alpha_{\odot}$), as seen in Figure \ref{fig:avg_rgb_temp}. The asteroseismic targets are in much stronger agreement with the isochrone at fainter magnitudes and have more scatter at the more evolved, brighter stars which can also be seen in the \mlt\ = 1.80 panel in Figure \ref{fig:NGC6791_all_isochrones_grid}. Additionally, we see there is a color offset relationship with bluer-than-expected, lower RGB stars and redder upper RGB stars. In Figure \ref{fig:NGC6791_all_isochrones_grid}, we see this offset in all models where the isochrone crosses the asteroseismic targets (\mlt\ $= 1.70 - 1.90$), which points to larger systematic offsets with model physics that can be improved in the future.

We also compare the average RGB temperature of the Gold asteroseismic targets to the isochrones in theoretical coordinates ($\log g$ vs $T_{\mathrm{eff}}$). Using the same method for calculating the difference between the observed targets and modeled points, we find the best fitting mixing length parameter to be $\alpha_\text{MLT} = 1.90$ ($1.10\alpha_{\odot}$). Figure \ref{fig:ngc6791_logg_isochrone} shows the asteroseismic targets on an $\alpha_\text{MLT} = 1.90$ isochrone with errors on both axes. In the left panel, the uncertainties in both $T_\text{eff}$ and $\log g$ have been multiplied by 2 for visualization purposes; however, the uncertainty in $\log g$ is still too small to be visible. The uncertainties in the residual plot on the right have not been modified. The errors are listed in Table \ref{tab:ngc6791_members}.
 
While we technically find a deviation between best-fitting mixing lengths depending on the coordinate system in which we fit---a best-fitting mixing length of 1.90 in the $\log g$-$T_\text{eff}$ plane and a best-fitting mixing length of 1.80 in the Gaia color-magnitude plane---it is important to note that, in the Gaia color-magnitude coordinate system, the goodness of fit of the $\alpha_\text{MLT} = 1.80$ isochrone and the goodness of fit of the $1.90$ isochrones are nearly identical. They are also much closer together (from Equation \ref{eqn:pythag_dist}, the Pythagorean distance for these two models is 0.30 vs 0.33, respectively) than the fits of other isochrones with adjacent mixing length parameters (typical variations of 10\% or more). To explore this, we ran an additional grid of models at a higher \mlt\ resolution of $0.05$, finding that an isochrone with $\alpha_\text{MLT} = 1.85$ was likewise functionally indistinguishable. We therefore assert that these MLT determinations are equivalent within the resolution of the grid considered.

By comparing the temperature offset of the asteroseismic targets to that of the predicted isochrone, we find a value of \mlt\ = 1.80 (1.04$\alpha_{\odot}$) or 1.90 (1.10$\alpha_{\odot}$) as the best fitting mixing length parameter, depending on the coordinate space.

\subsection{Metric 3: Curvature of the Red Giant Branch} \label{subsec:ngc6791_rgb_curve}

\begin{figure}
    \centering
    
    \begin{subfigure}[t]{0.7\linewidth}
        \includegraphics[width=1\linewidth]{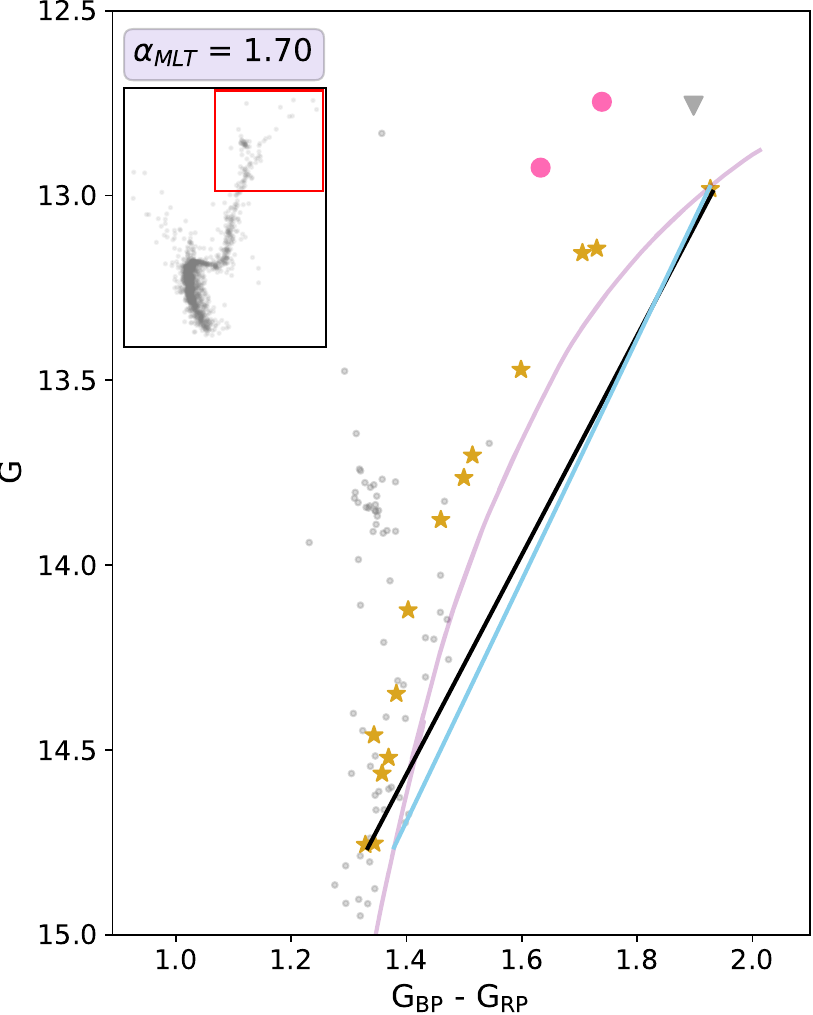}
        \caption{Comparing the curvature of only the Gold asteroseismic targets to the isochrone with $\alpha_\text{MLT} = 1.70$.}
        \label{subfig:rgb_gold_only}
    \end{subfigure}
    \medskip
    
    \begin{subfigure}[t]{0.7\linewidth}
        \includegraphics[width=1\linewidth]{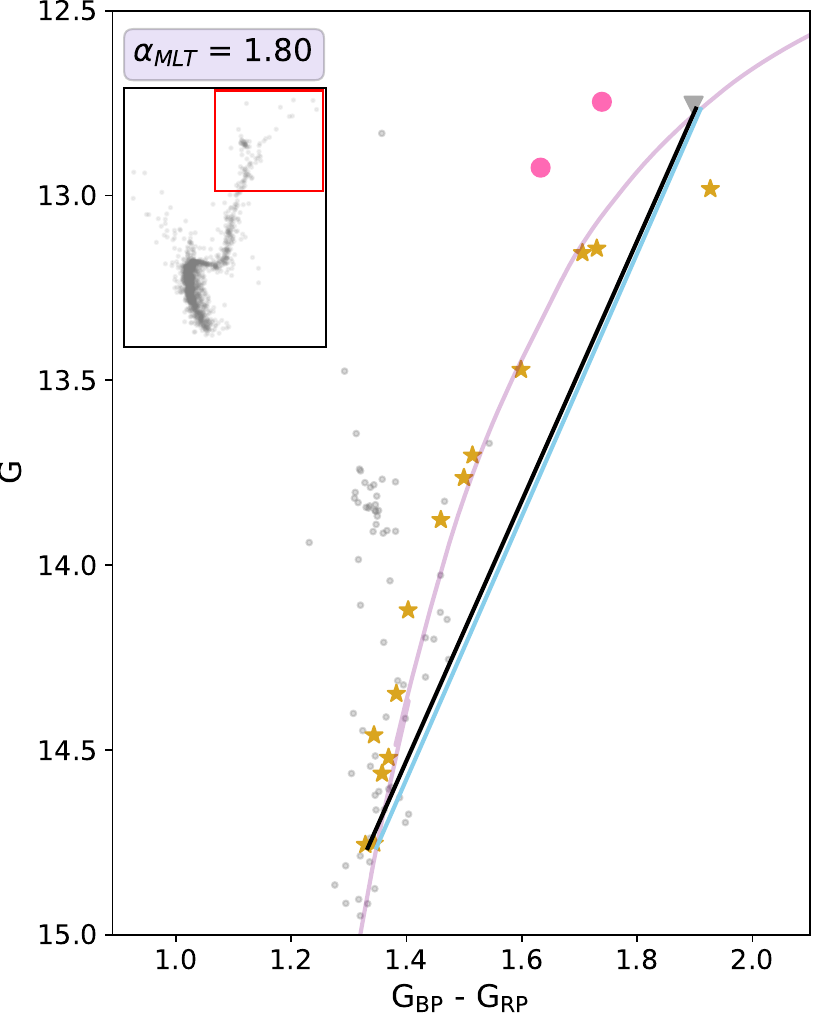}
        \caption{Comparing the curvature of the Gold and Silver asteroseismic targets to the isochrone with $\alpha_\text{MLT} = 1.80$.}
        \label{subfig:rgb_silver}
    \end{subfigure}
    
    \caption{Zoom in on RGB to show comparison between slope of NGC 6791 (black line) and the isochrone RGB (blue line) between dimmest and brightest asteroseismic targets. Figure \ref{subfig:rgb_gold_only} only includes the Gold targets while Figure \ref{subfig:rgb_silver} also includes the Silver target. The inset shows the location of the zoomed in RGB in relation to the whole cluster.}
    \label{fig:rgb_curvature}
\end{figure}

The RGB encodes useful information about clusters. For example, the slope is sensitive to metallicity \citep[e.g.,][]{ferraro_giant_1999, ferraro_new_2000} and is independent from age and distance \citep{valenti_red_2004}. Additionally, when modeling observations, a poorly fitting RGB can indicate problems with color transformations or the treatment of convection in models \citep{vandenberg_models_2000}. By comparing the curvature of isochrones to the precise seismic data on the RGB in a cluster for which the age and metallicity are extremely well-constrained, we can isolate $\alpha_\text{MLT}$ to explore how well different hypotheses for the treatment of convection fit the data.

To test the curvature of the RGB, we chose to focus on the range of the RGB between the dimmest and brightest asteroseismic targets (points C and D in Figure \ref{fig:morphology_definitions}). With the $G$ magnitudes of C and D, we found the corresponding point in the modeled isochrone and drew a straight line between the two. We then compared the location of the isochrone line to the cluster line using Equation \ref{eqn:pythag_dist}, replacing A and B with C and D. We also compared the difference between the slopes of both lines. The isochrone with the smallest Pythagorean distance and the smallest slope difference is the best \mlt. Additionally, we chose to compare the curvature of the RGB both including and excluding the Silver targets, which can be seen in Figure \ref{fig:rgb_curvature}, where the black line shows the slope of the cluster, and the blue line shows the slope of the isochrone, and the purple line is the isochrone.

Only factoring in the Gold targets (Figure \ref{subfig:rgb_gold_only}), we find that the best fitting mixing length parameter is $\alpha_{\rm MLT} = 1.70$. When including the Silver target (Figure \ref{subfig:rgb_silver}), we see that the best fitting isochrone has $\alpha_{\rm MLT} = 1.80$, with the isochrone fitting the majority of the seismic targets well.

\subsection{Metric 4: Offset from Most Luminous Target} \label{subsec:most_luminous}
\begin{figure}[b!t]
    \centering
    \includegraphics[width=0.9\linewidth]{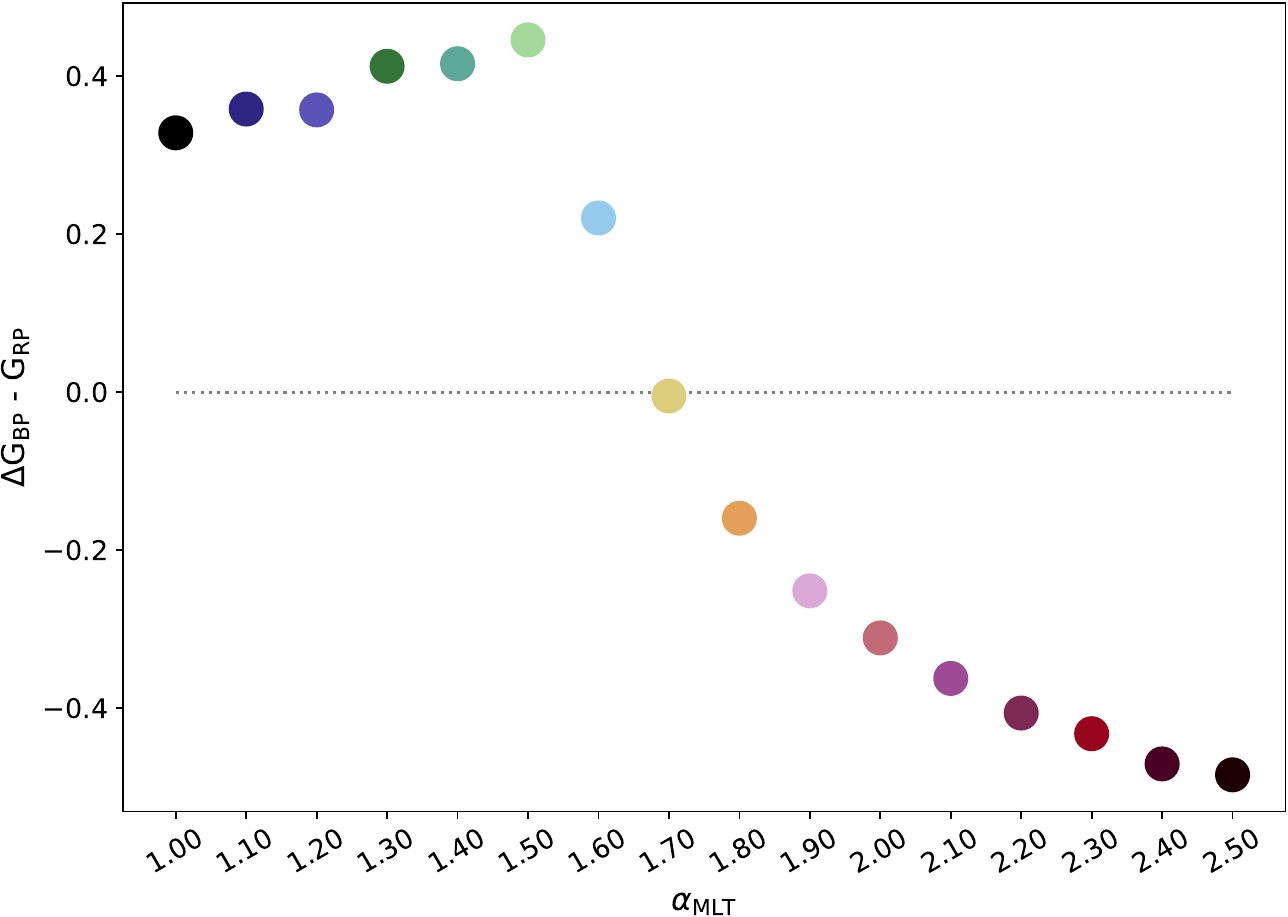}
    \caption{Comparison of the horizontal distance between most luminous Gold sample target (grey dotted line) and the corresponding point on each isochrone. The closest isochrone has a mixing length parameter of \mlt\ = 1.70 for NGC 6791.}
    \label{fig:most_luminous}
\end{figure}
\begin{figure*}[!]
    \centering
    \includegraphics[width=0.8\linewidth]{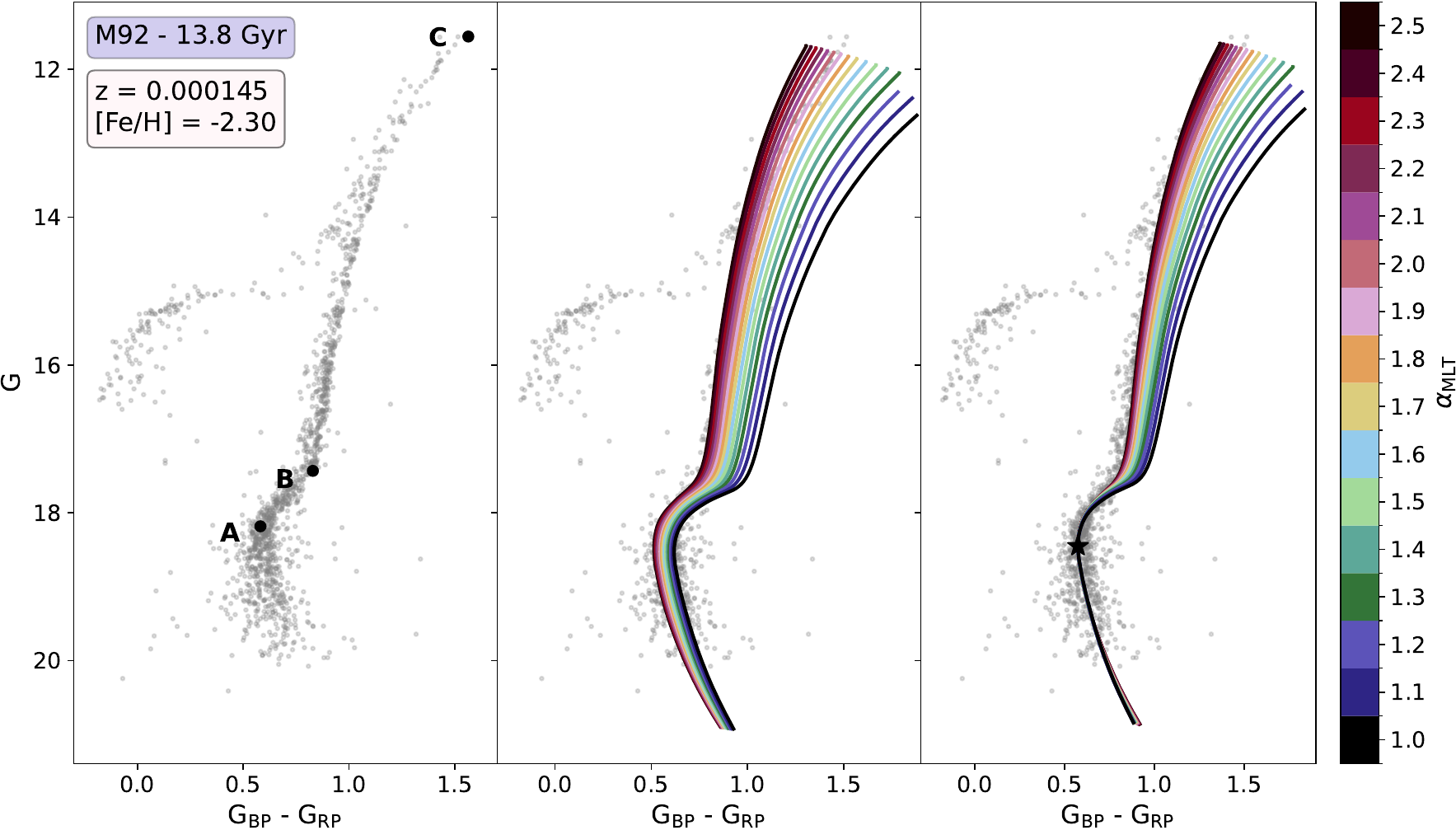}
    \caption{\textit{Left:} Same as Figure \ref{fig:morphology_definitions} but for M92. \textit{Center} \& \textit{Right:} Same as Figure \ref{fig:NGC6791_vs_all_isochrones} but for M92.}
    \label{fig:M92_vs_all_isochrones}
\end{figure*}
To prioritize the upper RGB into our analysis, we include a comparison of observations of the most luminous Gold-sample asteroseismic target to each isochrone. To do so, we found the location on the isochrone with the same $G$ magnitude as the most luminous star and determined the horizontal distance between them. The results of this can be seen in Figure \ref{fig:most_luminous}, where the isochrone with the smallest offset has a mixing length parameter of \mlt\ = 1.70, or 0.98$\alpha_{\odot}$, which agrees with the results of the RGB curvature for the Gold targets (Section \ref{subsec:ngc6791_rgb_curve}).

\section{Fitting Reference Cluster M92}\label{sec:M92_compare_to_obs}

Due to its rigorous observational characterization, M92 has been the subject of mixing length calibration analyses within the past few years in \citet{joyce_not_2018} and \citet{ying_absolute_2023}. We therefore perform our own analysis of this cluster using \mesa\ for the first time (all previous \mlt\ calibrations have been performed using DSEP).

We built a suite of customized isochrones with varying values of $\alpha_\text{MLT}$ for M92, each with age and metallicity tailored to the cluster, as discussed in Sections \ref{subsec:Model Grid} and \ref{subsec:isochrone building}.
We overlaid the computed 13.8 Gyr isochrones atop the \gaia\ DR3 data for M92. This is seen in Figure \ref{fig:M92_vs_all_isochrones}, where the center panel shows where the isochrones naturally fall on the cluster data and the right panel shows the isochrones after being aligned at the MSTO. To determine the MSTO point for M92, we first attempted the ridgeline method (discussed in Section \ref{sec:determinations_NGC6791}); however, when we applied the ridgeline method of \citet{simon-zsok_empirical_2025}, we found that it was not effective for this cluster given its morphology with fewer well--defined inflection points. We found it more effective to determine the cluster's MSTO by visual inspection.

Due to a lack of asteroseismic targets for M92, we have adopted only two of the best fit metrics: matching the SGB length and the RGB curvature. The left-hand panel of Figure \ref{fig:M92_vs_all_isochrones} shows three critical points in the morphology of M92, defined by visual inspection: point A marks the start of the SGB, point B shows the end of the SGB and the start of the RGB, and point C is the most luminous and red cluster star which represents the TRGB.

\subsection{Subgiant Branch Length}\label{subsec:M92_sgb}

\begin{figure}
    \centering
    \includegraphics[width=0.85\linewidth]{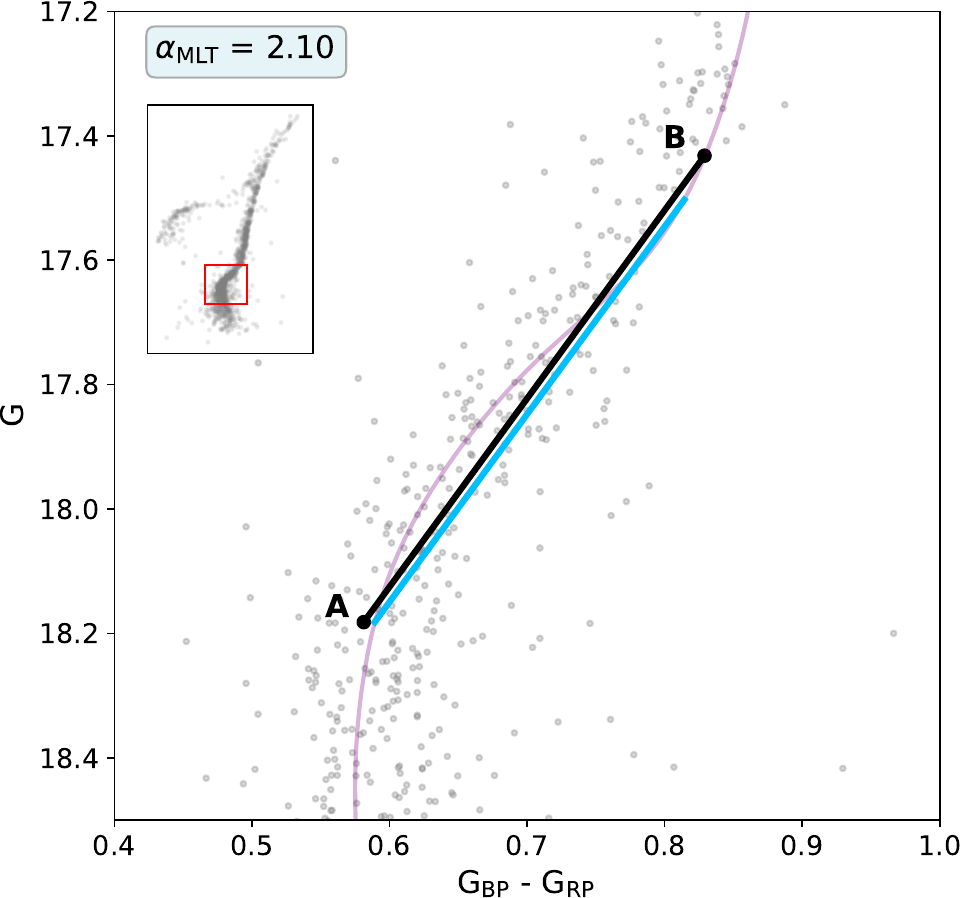}
    \caption{Same as Figure \ref{fig:subgiant_length_metric} but for M92 with $\alpha_\text{MLT} = 2.10$.}
    \label{fig:M92_SGB}
\end{figure}

Due to the old age and shape of M92, we are unable to use the same local maxima and minima determination method used to demarcate the SGB for NGC 6791. We found the position on the isochrone with the same \gaia\ $G$ magnitude as point A and took the corresponding value of $\text{G}_\text{BP}-\text{G}_\text{RP}$ to mark the start of the isochrone SGB. We defined the end of the SGB by mathematically finding the inflection point.

To determine the best fit, we compared the slope of the cluster's SGB to that of the isochrone's. We found that the best fitting isochrone to M92's SGB has $\alpha_{\rm MLT} = 2.10$ ($1.21\alpha_{\odot}$) as shown in Figure \ref{fig:M92_SGB}, where the subgiant branch is represented by a black line for the cluster and a blue line for the isochrone. The slope of the SGB almost perfectly matches the cluster's with the presence of a slight red color offset.

\subsection{Curvature of the Red Giant Branch}\label{subsec:M92_rgb}
\begin{figure}
    \centering
    \includegraphics[width=0.8\linewidth]{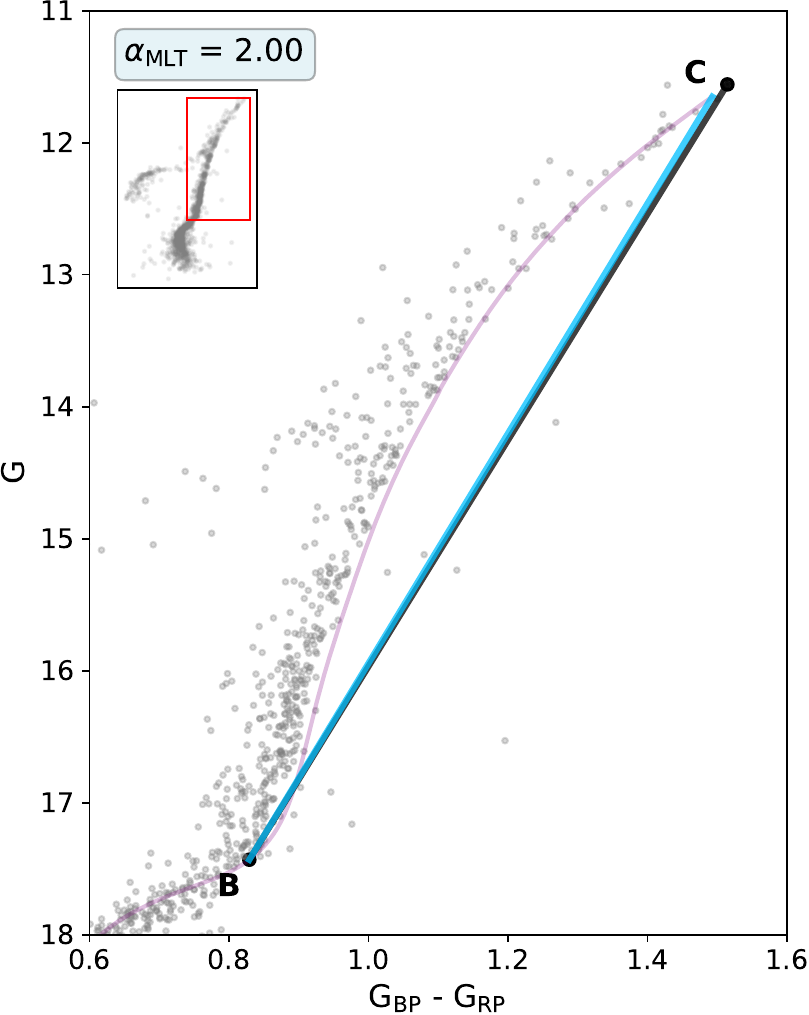}
    \caption{Same as Figure \ref{fig:rgb_curvature} but for M92 with $\alpha_\text{MLT} = 2.00$.}
    \label{fig:M92_rgb}
\end{figure}

The start of the isochrone's RGB is marked with the same isochrone inflection point representing the end of the SGB, as described in Section \ref{subsec:M92_sgb}, while the isochrone's TRGB is the most luminous point. Rather than align the isochrones at the MSTO for this analysis, we chose to align them at the RGB turnoff (point B) instead. \citet{joyce_not_2018} found that at a given age, the RGB is extremely sensitive to \mlt, thus by aligning each isochrone at the RGB turnoff, we are able to best analyze how the curvature of the RGB changes as a function of mixing length parameter independent of any morphological variations prior to this point.

To compare the isochrone with the \gaia\ data, we calculate the slope between points A and B and between the RGB of the isochrone. We calculate the difference between these two values and determine that the lines with the smallest divergence between, and is therefore the best fitting isochrone, has $\alpha_\text{MLT} = 2.00$ ($1.16\alpha_{\odot}$). Figure \ref{fig:M92_rgb} shows the cluster's RGB slope (black line) to the isochrones (blue line) for this \mlt.

\vspace{0.4cm} 

\section{Decreasing $\alpha_\text{MLT}$ Trend}

\begin{figure*}[tb]
    \centering
    \begin{subfigure}[t]{0.45\linewidth}
        \includegraphics[width=1\linewidth]{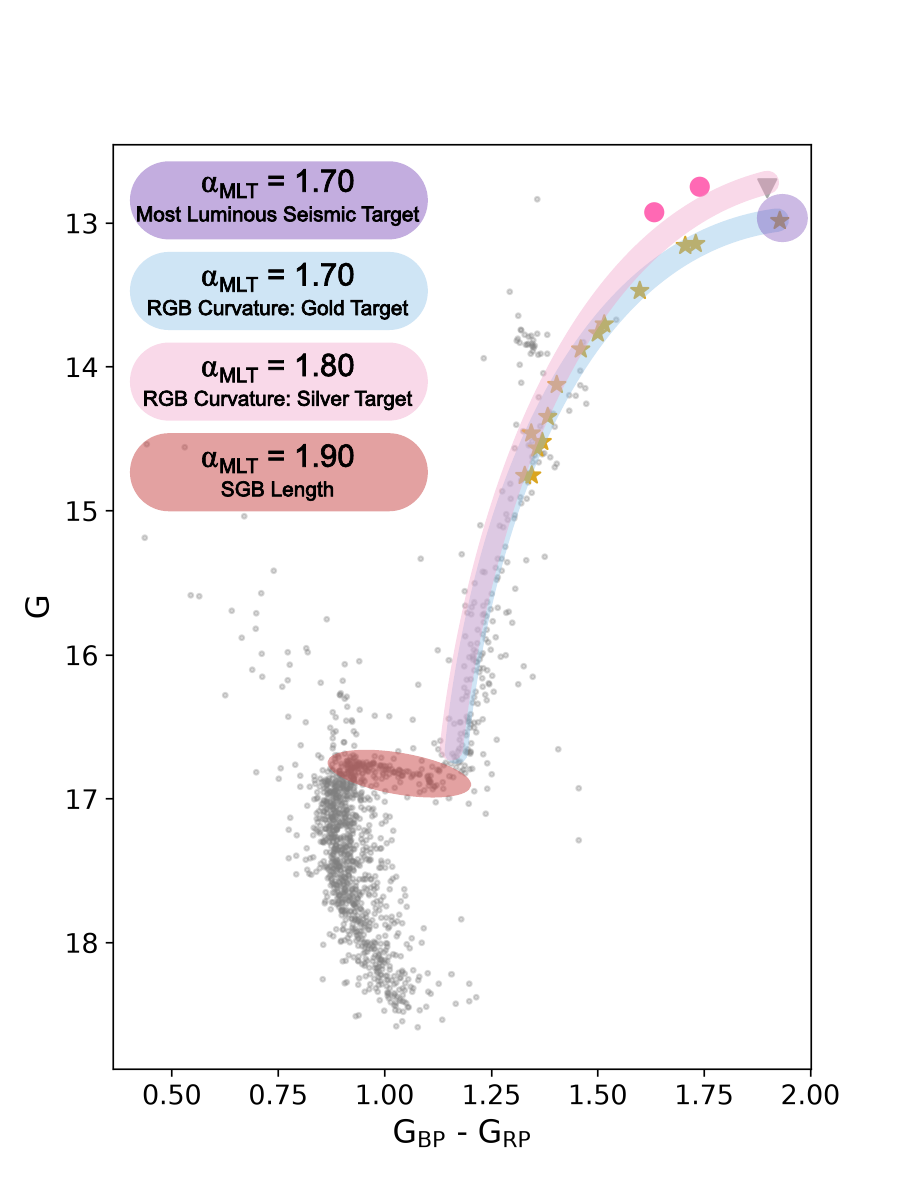}
        \caption{Best fits for NGC 6791. Does not include average RGB temperature (best fit $\alpha_\text{MLT} = 1.80$) or the $\log g-T_\text{eff}$ isochrone fit ($\alpha_\text{MLT} = 1.90$).}
        \label{subfig:NGC6791_summary}
    \end{subfigure}
    \begin{subfigure}[t]{0.45\linewidth}
        \includegraphics[width=1\linewidth]{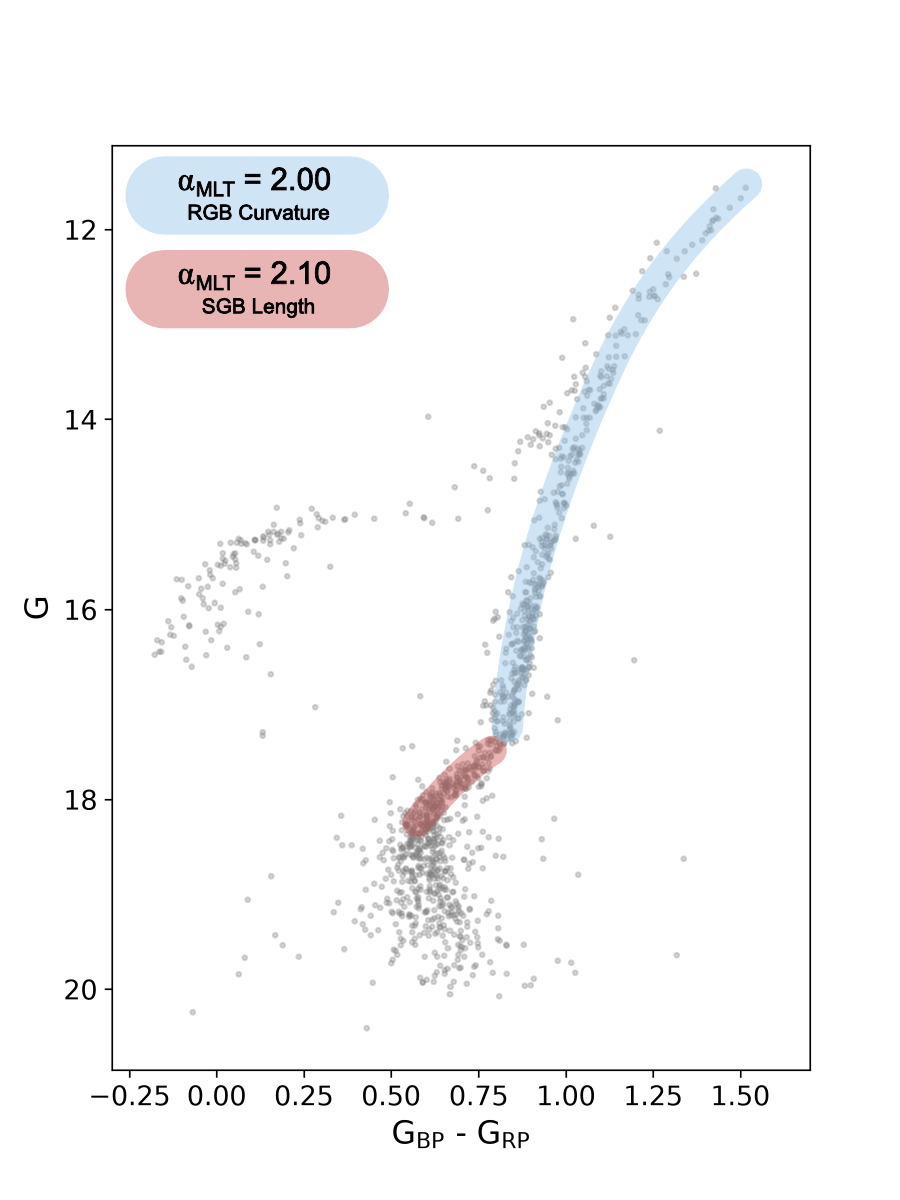}
        \caption{Best fits for M92.}
        \label{subfig:m92_summary}
    \end{subfigure}
    \caption{Visual summary of best fitting mixing length parameters from each metric (Table \ref{tab:best_fit_mlt})}
    \label{fig:best_fit_mlt}
\end{figure*}

For both clusters, we built a set of isochrones, each with a different, static value of \mlt\ adopted homogeneously in its constituent evolutionary tracks. With the availability of asteroseismic measurements of individual stars in the well-characterized, metal-rich open cluster NGC 6791, we were provided with a unique opportunity to test model physics in a highly constrained environment. Using the four different best fit metrics described in Sections \ref{subsec:ngc6791_sgb}-\ref{subsec:most_luminous}, we find that no single mixing length best fits NGC 6791 but rather a range between $1.70-1.90$, or $0.98-1.10 \alpha_{\odot}$ is better (results are shown in Table \ref{tab:best_fit_mlt} and Figure \ref{subfig:NGC6791_summary}). Finding that stars at earlier phases in evolution are better fit by larger mixing length parameters, while stars further along in evolution have lower best-fitting \mlt\ values. 

To compare M92 to our bespoke suite of isochrones, we used the best-fit metrics outlined in Sections \ref{subsec:M92_sgb} and \ref{subsec:M92_rgb}.
We find best-fit values of $2.00-2.10$, or $0.16-1.21 \alpha_{\odot}$, showing the same relationship of decreasing \mlt\ as stellar models evolve (results are shown in Table \ref{tab:best_fit_mlt} and Figure \ref{subfig:m92_summary}).

For both clusters, we determine that the best-fitting \mlt\ monotonically decreases with evolutionary stage. We find near-solar, decreasing values of \mlt\ for the metal-enriched open cluster NGC 6791 and super-solar, decreasing values of \mlt\ for the extremely metal-poor globular cluster M92. This trend does not agree with the \mlt--metallicity relationship from \citet{tayar_correlation_2017}, who claimed no trend with $\log g$---a proxy for evolutionary stage---in their sample of APOKASC-2 targets.

\subsection{Agreement with 3D Results}\label{sec:comp_to_3D}

Exploring mixing length parameter calibration is not novel \citep[e.g.][]{salaris_new_1996, bonaca_calibrating_2012, tayar_correlation_2017, joyce_not_2018}. Though the scope of this paper and much of the literature has focused on \mlt\ in one dimension, a handful of studies have attempted to consider the mixing length in 3D contexts, such as \citet{trampedach_improvements_2014} and \citet{magic_stagger-grid_2015}.

By matching 1D envelope models to 3D radiation-coupled hydrodynamic (RHD) simulations of convection,   \citet{trampedach_improvements_2014} found that stars on the pre-main sequence would have evolved with higher \mlt\ than those on the main sequence and beyond. Shortly after, \citet{magic_stagger-grid_2015} used realistic 3D RHD simulations to calibrate the mixing length parameter using a 1D MLT stellar atmosphere code. They found that the mixing length parameter decreases as $\log g$ decreases and as $T_\text{eff}$ and [Fe/H] increase. When looking at the \mlt\ variation across stellar evolution, deviations from the solar mixing length parameter up to $\pm20\%$ were found. These results are visualized in Figure 9 of \citet{tayar_correlation_2017}. For \citet{magic_stagger-grid_2015} targets, we see a correlation of decreasing mixing length with decreasing $\log g$, and thus evolutionary stage.

To our knowledge, there have been no further studies which conclude that the mixing length parameter decreases with evolutionary stage, though it is thought that \mlt\ should not be constant through evolution \citep[e.g.,][]{song_impact_2020, reyes_isochrone_2024}. Additionally, there have been no others using one dimensional codes that have revealed this relationship, and thus that reproduce the findings of the 3D RHD results from \citet{magic_stagger-grid_2015}.

\subsection{Persistence Across Variations in Atmospheric Boundary Conditions}\label{sec:atm_BCs}

As discussed in Section \ref{sec:Intro}, there are a number of physical assumptions that can impact the resulting stellar tracks; the choice of atmospheric boundary conditions is among them. \citet{choi_red_2018} found that the choice of atmospheric boundary condition can impart uncertainties up to $\sim100$ K on the RGB. To explore how changing the boundary condition can impact the shape of the RGB, we recreated our isochrones for both clusters using a gray model atmosphere that follows the Eddington \textit{T}$-\tau$ relation.

Further, to test how the optical depth, $\tau$, at which the boundary conditions are applied can impact the stellar tracks, we built three sets of isochrones with gray model atmospheres at depths $\tau = 1/3,\ 2/3$,\ and $1$. We then ran associated solar calibrations using the same method as described in Section \ref{subsec:solar_calibration}, finding that calibrated under a gray model atmosphere patched at $\tau=2/3$, $\alpha_{\text{MLT},\odot} = 1.98995$, increasing by $\sim$15\% compared to when using PHOENIX, corroborating the findings of \citet{joyce_not_2018}. We then used the same best fit metrics outlined in Sections \ref{sec:determinations_NGC6791} and \ref{sec:M92_compare_to_obs} for the NGC 6791 and M92 data, respectively. We find that for both clusters, while the global value of \mlt\ changes (which is to be expected when altering model physics), regardless of the depth applied or the boundary condition used, the value of \mlt\ always decreases monotonically as a function of evolutionary phase.

\begin{deluxetable}{c|cccc}
    \tablecaption{Best fitting mixing length parameters for each metric and compared to the solar calibrated value of $\alpha_{\odot} = 1.72928$. Visualized on cluster data in Figure \ref{fig:best_fit_mlt}. \label{tab:best_fit_mlt} }
    
        \tablehead{
        \colhead{\textbf{Cluster}} &
        \colhead{\textbf{Method}} &
        \colhead{\bm{$\alpha_{\rm MLT}$}} &
        \colhead{\bm{$\alpha_{\rm MLT}/\alpha_{\odot}$}}
        }

        \startdata
        NGC 6791 & $\log g-T_{\rm eff}$ isochrone fit & 1.90 & 1.10 \\
        & SGB length & 1.90 & 1.10 \\
        & Average RGB temperature & 1.80 & 1.04 \\
        & RGB curvature: Silver & 1.80 & 1.04 \\
        & RGB curvature: Gold & 1.70 & 0.98 \\
        & Most luminous seismic target & 1.70 & 0.98 \\
        \hline
        M92 & SGB length & 2.10 & 1.21 \\
        & RGB curvature & 2.00 & 1.16
        \enddata 
        
\end{deluxetable}

\section{Conclusions} \label{sec:conclusion}

We have demonstrated that the best-fitting mixing length decreases monotonically with evolutionary stage, from the MSTO to late on the red giant branch, for both a metal-rich open cluster and an old, metal-poor globular cluster.
Interestingly, this trend persists across very
different metallicities: [Fe/H] = $+0.31$ and $-2.30$. Importantly, we have also shown that the degree of variation among \mlt\ values fit to different regions of the same object---a variation of 18\% in the case of NGC 6791---is comparable to the roughly 20\% spread in \mlt\ determinations for, e.g., members of the near-solar-analog binary alpha Centauri A \& B \citep{joyce_classically_2018}. 
Similarly, \citet{tayar_correlation_2017} found a $\sim$40\% spread in \mlt\ for $\sim$3000 APOKASC-2 red giants, and we find an 18\% spread for stars of a single cluster at different evolutionary phases.

While our finding that M92's best-fitting mixing length is super-solar is in tension with previous determination efforts---for both M92 itself and for stars and objects with similarly low metallicities---it should be emphasized that this is the first such study to run this calibration with \mesa, which differs significantly from, e.g., DSEP, including in that it is not optimized for low-mass modeling.\footnote{Whereas DSEP is optimized for low-mass stellar modeling, \mesa\ is not, and we should therefore trust the absolute \mlt\ values returned by DSEP for clusters of low-mass stars over \mesa's. However, the absolute values do not impact the relative \mlt\ fits in different regions of M92.} Further, the solar calibration that provides the $\alpha_\text{MLT}$ normalization was performed by allowing variation in the helium enrichment quantity $\frac{\Delta Y}{ \Delta Z}$, whereas our isochrones adopt a fixed value (see Table \ref{tab:cluster_properties}). It is therefore most informative to compare relative values of the mixing length in this analysis.

This $18\%$ spread in \mlt\ for NGC 6791 is larger than the degree of variation we typically see across all mixing length determinations for a single object in the literature. However, our findings are consistent with the fact that there is no evidence that \mlt\ should remain static throughout a evolutionary model \citep{song_impact_2020, reyes_isochrone_2024}. Using 3D hydrodynamical simulations, \citet{magic_stagger-grid_2015} found \mlt\ to decrease as $\log g$ decreases and $T_\text{eff}$ and [Fe/H] increase, a trend that we replicate in 1D using \mesa. Additionally, they found that \mlt\ can change up to $20\%$ with stellar mass, which is consistent with the $18\%$ spread in \mlt\ we find for NGC 6791.

By building a bespoke grid of isochrones with varying mixing length parameters, we have demonstrated that:
\begin{enumerate}
    \item \mlt\ depends not just on mass and metallicity, but also decreases monotonically as a function of evolutionary stage;
    \item this trend persists across highly divergent metallicities ([Fe/H] = $+0.31$ and $-2.30$) for both an open and globular cluster, respectively;
    \item it also persists when changing both the atmospheric boundary condition used in the stellar models, and the optical depth at which these boundary conditions are applied;
    \item the 18\% variation found within fits to one single object (NGC 6791) is comparable to variation across the literature in \mlt\ fits for other non-solar targets, calling single-valued \mlt\ fits to clusters into question; and
    \item our findings reproduce the trend of decreasing \mlt\ as stars evolve (with increasing $T_\text{eff}$ and decreasing $\log g$) from \citet{magic_stagger-grid_2015} in 1D.
\end{enumerate}

\section{Future Work}
\label{sec:future_work}
In this paper, we have not only reinforced the established conclusion that non--solar mixing length parameters are required to accurately model star clusters, but demonstrated explicitly that the best-fitting mixing length is a function of evolutionary stage within single populations.
Astronomers have explored \mlt\ as a free parameter and concluded it is necessary to explore it in conjunction with mass and metallicity among other physical properties. However, there are currently no open source databases of isochrones that include variations in static \mlt\ nor publicly available models that change \mlt\ adaptively, as our results demonstrate is required. Thus it is prudent to create an \mlt--adaptive database of isochrones in order to improve stellar age determinations and to learn more about the internal processes inside of stellar populations.

 In the forthcoming era of stellar science driven by Roman \citep{akeson_wide_2019}, LSST \citep{ivezic_lsst_2019}, and Plato \citep{rauer_plato_2014} (see Table 1 of \citealt{miller_custom_2026} for a summary of stellar science goals of upcoming missions) and with access to the better data quality and high precision asteroseismic observations in crowded fields enabled by these missions, having an open source database of isochrones with varying mixing lengths is crucial to better characterize these clusters.

\begin{acknowledgements}

    Computations were performed using the University of Wyoming (UW) Advance Research Computing Center MedicineBow HPC, a UW managed computational resource available to UW researchers including faculty, staff, students, and collaborators.

    This research has made use of public auxiliary data provided by ESA/Gaia/DPAC/CU5 and prepared by Carine Babusiaux. Additionally, it has made use of data from the European Space Agency (ESA) mission {\it Gaia} (\url{https://www.cosmos.esa.int/gaia}), processed by the {\it Gaia} Data Processing and Analysis Consortium (DPAC, \url{https://www.cosmos.esa.int/web/gaia/dpac/consortium}). Funding for the DPAC has been provided by national institutions, in particular the institutions participating in the {\it Gaia} Multilateral Agreement.

\end{acknowledgements}

\facilities{ARCC \citep{advanced_research_computing_center_arcc_medicinebow_2018}, \textit{Gaia}, SDSS, and Kepler.}
\software{TOPCAT \citep{taylor_topcat_2005}, Matplotlib \citep{hunter_matplotlib_2007}, pandas \citep{mckinney_data_2010}, and numpy \citep{harris_array_2020}.}

\bibliography{HolyGrailSideQuest}{}
\bibliographystyle{aasjournalv7}

\appendix
\section{\textit{Gaia} Query Parameters} \label{app:query_params}

\subsection{NGC 6791}\label{app:query_params_NGC6791}

After querying \gaia\ by 3 times the half-light radius, we took the following steps to ensure the best quality data from our cluster members: 

\renewcommand\labelitemi{\tiny$\bullet$}

\begin{itemize}
    \item {\tt phot\_bp\_mean\_mag} $\bm{< 20.3}$. For faint red sources, the mean G$_{BP}$ is overestimated due to an implemented lower flux limit as described in Figure 27 and Section 8.1 of \citet{riello_gaia_2021}.
    \item {\tt parallax} $\bm{\pm\ 0.1}$ to remove any stars that are not at the same distance as the clusters. For NGC 6791 it was {\tt parallax} between 0.09 and 0.29.
    \item {\tt parallax\_error} $\bm{< 0.15}$ \citep{cantat-gaudin_gaia_2018}
    \item {\tt pmra} between -1 and 0 and {\tt pmdec} between -3 and -1.7. These selections were made by potting queried stars on a {\tt pmra} vs {\tt pmdec} plot and visually selecting the upper and lower limits on the clump location. The cluster before {\tt pmra} and {\tt pmdec} cuts with a clear distinction between cluster and field stars can be seen in Figure \ref{fig:app:pmra_vs_pmdec}. We can make the assumption that objects not at similar proper motion as the observed clump are not located in the cluster.
    \item {\tt pmra\_error} and {\tt pmdec\_error} $\bm{< 0.3}$ \citep{cantat-gaudin_gaia_2018}
    \item {\tt ruwe} $< \textbf{1.15}$ \citep{vasiliev_gaia_2021}. RUWE is the renormalized unit weight error for astrometry. Values above 1.4 indicate binary stars and stars in younger open clusters ($<$100 Myr) \citep{babusiaux_gaia_2023}
    \item {\tt astrometric\_excess\_noise\_sig} $\bm{< 2}$. Gives the significance of excess source noise for each observation, and is considered statistically significant above 2 \citep{vasiliev_gaia_2021, lindegren_gaia_2021}. Can also be a binary star indicator.
    \item {\tt ipd\_gof\_harmonic\_amplitude} $ \bm{<\ 0.1}$. Indicates asymmetry in images \citep{fabricius_gaia_2021}.\footnote{According to \citet{fabricius_gaia_2021}, if a source's {\tt ipd\_gof\_harmonic\_amplitude} $>\ 0.1$ and {\tt ruwe} $>\ 1.4$, it is indicative of resolved doubles.} 
    \item {\tt ipd\_frac\_multi\_peak} $\bm{< 2}$, indicating the presence of resolved, close pairs \citep{fabricius_gaia_2021}.
    \item {\tt visibility\_periods\_used} $\bm{> 10}$ \citep{riello_gaia_2021}. Visibility periods are epochs of observations with no a maximum time gap between observations of 4 days. In order for proper motion and parallaxes to be published, there must be at least 9 visibility periods, so this parameter ensures all sources we choose will have proper motion and parallax \citep{fabricius_gaia_2021}.
\begin{figure}[t]
    \centering
    \includegraphics[width=0.9\linewidth]{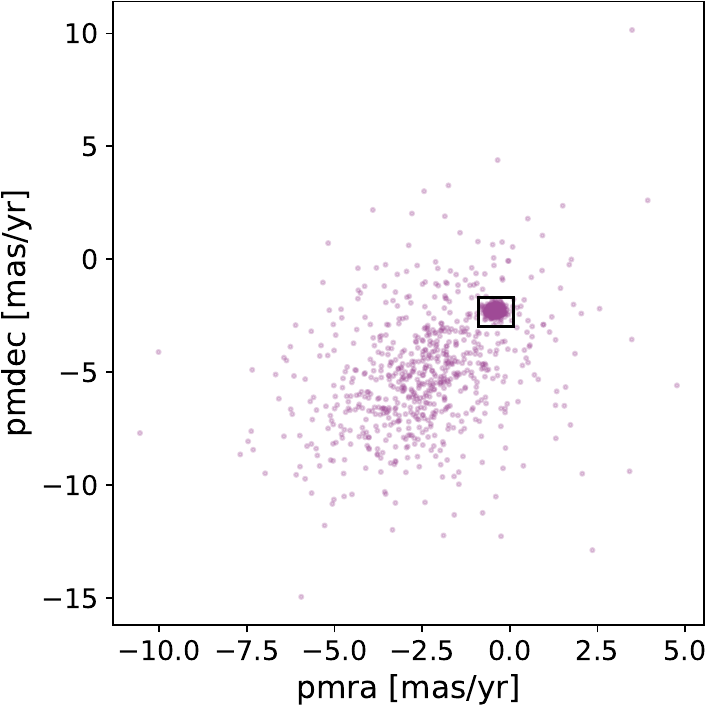}
    \caption{NGC 6791 cluster members before making cuts for proper motion. There is a clear group of objects which we know to be stars in the cluster, highlighted by the black box.}
    \label{fig:app:pmra_vs_pmdec}
\end{figure}
    \item {\tt phot\_bp\_rp\_excess\_factor} limited via Equation 6 and Table 2 from \citet{riello_gaia_2021}. This is used to ensure sources have consistent photometry in the $G$-band, BP, and RP. It says

    \begin{equation*}
        C^{*} = C - f(\mathrm{G_{BP}} - \mathrm{G_{RP}}),
    \end{equation*} 
    where C = {\tt phot\_bp\_rp\_excess\_factor} and $f(\mathrm{G_{BP}} - \mathrm{G_{RP}})$ is

    \begin{equation*}
        \begin{aligned}
            f(\mathrm{G_{BP}} - \mathrm{G_{RP}}) & = 1.154360\ \\
            & + 0.033772(\mathrm{G_{BP}} - \mathrm{G_{RP}}) \\
            & +\ 0.032277(\mathrm{G_{BP}} - \mathrm{G_{RP}})^{2},
        \end{aligned} 
    \end{equation*}
    for $(\mathrm{G_{BP}} - \mathrm{G_{RP}}) <\ 0.5$, and 
    \begin{equation*}
        \begin{aligned}
            f(\mathrm{G_{BP}} - \mathrm{G_{RP}}) & = 1.162004\ \\
            & + 0.011464(\mathrm{G_{BP}} - \mathrm{G_{RP}}) \\
            & +\ 0.049255(\mathrm{G_{BP}} - \mathrm{G_{RP}})^{2} \\ 
            & -\ 0.005879(\mathrm{G_{BP}} - \mathrm{G_{RP}})^{3},
        \end{aligned} 
    \end{equation*}
    for $0.5\ <\ (\mathrm{G_{BP}} - \mathrm{G_{RP}})\ < 4.0$.
\end{itemize}

\subsection{M92}\label{app:query_params_M92}

After querying \gaia\ with a circle centered around M92 with a radius approximately three times the half-light radius, we take the following steps to constrain the cluster members for the same reasons as in Appendix \ref{app:query_params_NGC6791}:

\begin{itemize}
    \item \texttt{phot\_bp\_mean\_mag} $\bm{< 20.3}$
    \item \texttt{parallax} between \textbf{0.0012} and \textbf{0.212}
    \item \texttt{parallax\_error} $\bm{< 0.15}$ 
    \item \texttt{pmra} between \textbf{-5.5} and \textbf{-4.5} and \texttt{pmdec} between \textbf{-1.5} and \textbf{0.5}

\end{itemize}

\section{Theory to Magnitude}
\label{app_subsec:theory_to_mags}

To compare our theoretical isochrones to observational \textit{Gaia} data, we transformed luminosity to \textit{Gaia} $G$-band magnitudes. We first calculated the absolute bolometric magnitude using the standard conversion
\begin{equation} \label{eqn:L_to_Mbol}
    M_{bol} = M_{bol,\odot} - 2.5\log\Big(\frac{L}{L_{\odot}}\Big),
\end{equation}
where $M_{bol,\odot}$ = 4.75 \citep{allen_astrophysical_1976, torres_use_2010}. With the absolute bolometric magnitude, we found the absolute magnitude in the $G$ band using the relationship
\begin{equation} \label{eqn:Mbol_to_Mabs}
    M_{G} = M_{bol} - BC_{G}(T_\mathrm{eff}, \log g).
\end{equation}
A description of the bolometric correction calculations and choice of dust extinction ratio is described in Section \ref{subsec:theory_to_mags}.

Using the distance modulus rearranged, we solved for the apparent magnitude in the $G$ band,
\begin{equation} \label{eqn:Mabs_to_mapp}
    m_{G} = 5\log\Big(\frac{d}{10 pc}\Big) + M_{G},
\end{equation}
where $d$ is the distance to the cluster in parsecs. This is the final step converting luminosity to apparent $G$.

To convert from $T_\mathrm{eff}$ to magnitude, we can use the Stefan-Boltzmann law and the flux--luminosity relationship to rewrite Equation \ref{eqn:L_to_Mbol} as

\begin{equation} \label{eqn:Teff_to_Mbol}
    M_{bol} = M_{bol,\odot} - 2.5\log\Big(\frac{d^2T_\mathrm{eff}^4}{d_{\odot}^2T_\mathrm{eff, \odot}^4}\Big).
\end{equation}
With $M_{bol}$ as a function of $T_\mathrm{eff}$, we then followed Equations \ref{eqn:Mbol_to_Mabs} and \ref{eqn:Mabs_to_mapp} to find the apparent magnitude for each band. Lastly, to find $G_{BP} - G_{RP}$, we took our results from Equation \ref{eqn:Mabs_to_mapp} and found the difference between them.

\section{NGC 6791 Asteroseismic Targets}

Table \ref{tab:ngc6791_members} provides a subset of Table 4 from \citet{pinsonneault_apokasc-3_2025} and Table 2 from \citet{tayar_star-crossed_2025}, including only the relevant NGC 6791 asteroseismic targets.

\begin{deluxetable*}{ccccccccc}[!]
    \tablecaption{Asteroseismic cluster members of NGC 6791 from \citet{tayar_star-crossed_2025} \& \citet{pinsonneault_apokasc-3_2025}.\label{tab:ngc6791_members}}

    \tablehead{
    \colhead{\textbf{KIC ID}} &
    \colhead{\textbf{\textit{Gaia} DR3}} &
    \colhead{\textbf{Mass}} &
    \colhead{\textbf{$\log g$}} &
    \colhead{\textbf{[Fe/H]}} &
    \colhead{\textbf{[}\bm{$\alpha$}\textbf{/Fe]}} &
    \colhead{\bm{$T_{\mathrm{eff}}$}} &
    \colhead{\textbf{Age}} &
    \colhead{\textbf{Classification}} 
    \\
    \colhead{} &
    \colhead{} &
    \colhead{(M\sun)} &
    \colhead{} &
    \colhead{} &
    \colhead{} &
    \colhead{(K)} &
    \colhead{(Gyr)} &
    \colhead{}
    }

    \startdata
        2436209 & 2051286795872609920 & $1.18 \pm 0.05$ & $2.65 \pm 0.005$ & 0.324 & 0.012 & $4424 \pm 55$ & 7.86 & Gold \\
        2436332	& 2051288135902691840 & $1.14 \pm 0.05$ & $2.34 \pm 0.004$ & 0.357 & 0.041 & $4298 \pm 46$ & 9.05 & Gold \\
        2569204	& 2051294251936167424 & $1.27 \pm 0.05$ & $1.62 \pm 0.004$ & 0.281 & 0.025 & $3926 \pm 41$ & 6.02 & Gold \\
        2436540 & 2051287311268706432 & $1.14 \pm 0.05$ & $2.66 \pm 0.005$ & 0.309 & 0.007 & $4421 \pm 55$ & 8.75 & Gold \\
        2569360 & 2051294045777739520 & $1.15 \pm 0.04$ & $2.22 \pm 0.004$ & 0.270 & 0.023 & $4260 \pm 46$ & 8.49 & Gold \\
        2436688 & 2051286933311576192 & $1.09 \pm 0.04$ & $2.77 \pm 0.005$ & 0.316 & 0.024 & $4471 \pm 56$ & 10.56 & Gold \\
        2436814 & 2051293221143725952 & $1.05 \pm 0.04$ & $2.28 \pm 0.005$ & 0.291 & 0.038 & $4299 \pm 47$ & 12.01 & Gold \\
        2436884 & 2051293152424246912 & $1.17 \pm 0.04$ & $1.83 \pm 0.004$ & 0.294 & 0.025 & $4021 \pm 44$ & 8.05 & Gold \\
        2436900 & 2051287070750544128 & $1.13 \pm 0.04$ & $2.44 \pm 0.005$ & 0.257 & 0.044 & $4363 \pm 53$ & 8.84 & Gold \\
        2569618 & 2051294114497261952 & 1.17 $\pm$ 0.05 & 2.64 $\pm$ 0.005 & 0.300 & 0.036 & 4427 $\pm$ 54 & 8.17 & Gold \\
        2437340 & 2051292877546340864 & 1.17 $\pm$ 0.05 & 1.80 $\pm$ 0.004 & 0.298 & 0.032 & 4020 $\pm$ 43 & 7.94 & Gold \\
        2569935 & 2051293319920668160 & 1.04 $\pm$ 0.04 & 1.60 $\pm$ 0.004 & 0.320 & 0.035 & 4006 $\pm$ 34 & 12.01 & Gold \\
        2437496 & 2051293118064539520 & 1.11 $\pm$ 0.17 & 1.52 $\pm$ 0.01 & 0.289 & 0.030 & 3898 $\pm$ 35 & 9.41 & Silver \\
        2437507 & 2051105135938204160 & 1.23 $\pm$ 0.05 & 2.20 $\pm$ 0.004 & 0.308 & 0.033 & 4243 $\pm$ 42 & 6.84 & Gold \\
        2437653 & 2051293118064542976 & 1.18 $\pm$ 0.05 & 2.77 $\pm$ 0.005 & 0.319 & 0.056 & 4534 $\pm$ 51 & 7.91 & Gold \\
        2437965 & 2051105754413616000 & 0.84 $\pm$ 0.03 & 1.74 $\pm$ 0.004 & 0.311 & 0.029 & 4107 $\pm$ 39 & 25.81 & Gold \\
        2570518 & 2051293599100902016 & 1.13 $\pm$ 0.05 & 2.56 $\pm$ 0.004 & 0.284 & 0.054 & 4456 $\pm$ 47 & 9.15 & Gold \\
        2438333 & 2051104276944822912 & 1.14 $\pm$ 0.04 & 2.68 $\pm$ 0.004 & 0.319 & 0.074 & 4504 $\pm$ 53 & 8.74 & Gold \\
        2708270 & 2051296760197357184 & 1.15 $\pm$ 0.05 & 2.03 $\pm$ 0.004 & 0.287 & 0.035 & 4165 $\pm$ 36 & 8.37 & Gold \\
        2571093 & 2051107609839671808 & 0.93 $\pm$ 0.26 & 1.16 $\pm$ 0.01 & 0.300 & 0.025 & 3724 $\pm$ 31 & 17.66 & Silver \\     
    \enddata
\end{deluxetable*}

\end{document}